\documentclass[apj, iop, numberedappendix]{emulateapj}

\usepackage[utf8]{inputenc}
\usepackage[T1]{fontenc}
 
\usepackage[backref,breaklinks,colorlinks,citecolor=blue]{hyperref}

\usepackage{ulem}

\shorttitle{The z=5.7 IGM opacity-density relation}
\shortauthors{Christenson et al.}

\usepackage{graphicx}
\usepackage{amssymb}
\usepackage{amsmath}
\usepackage{cleveref}

\usepackage{xcolor}
\usepackage{ulem}

\usepackage{natbib}
\usepackage{tabularx} 
\usepackage{multirow} 

\begin{document}

\newcommand{\kms}{{km\ s$^{-1}$}}
\newcommand{\flux}{ergs\ s$^{-1}$\ cm$^{-2}$}
\newcommand{\lum}{ergs\ s$^{-1}$}
\newcommand{\lya}{Ly$\alpha$}
\newcommand{\lyb}{Ly$\beta$}
\newcommand{\teff}{\tau_{\rm eff}}

\newcommand{\numLAES}{{404}}

\def\gdb#1{{\color{orange}{#1}\color{black}}}
\def\gdbcomment#1{{\color{orange}{\bf [#1]}\color{black}}}
\def\hmc#1{\textbf{{\color{violet}{#1}\color{black}}}}
\def\cut#1{{\color{orange}\sout{#1}\color{black}}}



\title{The relationship between IGM \lya\ opacity and galaxy density near the end of reionization}

\author{Holly M. Christenson\altaffilmark{1}, George D. Becker\altaffilmark{1}, Anson D'Aloisio\altaffilmark{1},Frederick B. Davies\altaffilmark{3}, Yongda Zhu\altaffilmark{1}, Elisa Boera\altaffilmark{4}, Fahad Nasir\altaffilmark{3}, Steven R. Furlanetto\altaffilmark{2}, Matthew A. Malkan\altaffilmark{2}}

\altaffiltext{1}{Department of Physics  Astronomy, University of California, Riverside, CA 92521, USA; hchri004@ucr.edu}
\altaffiltext{2}{Department of Physics and Astronomy, University of California, Los Angeles, CA 90095, USA}
\altaffiltext{3}{Max Planck Institut f$\ddot{\rm{u}}$r Astronomie, Heidelberg, Germany}
\altaffiltext{4}{INAF — Osservatorio Astronomico di Trieste, Trieste, Italy}

\begin{abstract}
Observed scatter in the \lya\ opacity of quasar sightlines at $z<6$ has motivated measurements of the correlation between \lya\ opacity and galaxy density, as models that predict this scatter make strong and sometimes opposite predictions for how they should be related. Our previous work associated two highly opaque \lya\ troughs at $z\sim5.7$ with a deficit of Lyman-$\alpha$ emitting galaxies (LAEs). In this work, we survey two of the most highly transmissive lines of sight at this redshift, towards the $z=6.02$ quasar SDSS J1306+0356 and the $z=6.17$ quasar PSO J359-06.  We find that both fields are underdense in LAEs within 10 $h^{-1}$ Mpc of the quasar sightline, somewhat less extensive than underdensities associated with \lya\ troughs.  We combine our observations with three additional fields from the literature, and find that while fields with extreme opacities are generally underdense, moderate opacities span a wider density range. The results at high opacities are consistent with models that invoke UV background fluctuations and/or late reionization to explain the observed scatter in IGM \lya\ opacities. There is tension at low opacities, however, as the models tend to associate lower IGM \lya\ opacities with higher densities.  Although the number of fields surveyed is still small, the low-opacity results may support a scenario in which the ionizing background in low-density regions increases more rapidly than some models suggest after becoming ionized.  Elevated gas temperatures from recent reionization may also be making these regions more transparent.
\end{abstract}



\section{Introduction}\label{sec:intro}
Understanding when and how cosmic hydrogen reionization proceeded is of great interest for several reasons. First, the timing and duration of reionization have implications for our understanding of the first luminous sources. Second, our understanding of the physical state of the IGM is important context for high-redshift observations that are affected by absorption by intervening material. Lastly, reionization functions as a test of our dark matter and galaxy formation models, which must produce sources consistent with reionization constraints. There are two primary open questions that current reionization studies are attempting to address: the timing of reionization, including when it ended, and what the main sources of ionizing photons are (see \citealt{wise19} for a review).

A number of observations suggest that much of reionization took place between z$\sim6-8$.  Damping wings in quasar spectra at $z\geq7$ suggest that the IGM is still substantially neutral at those redshifts \citep{mortlock11,greig17,banados18, greig19,davies18b,wang20}. Galaxy surveys infer that a large portion of the universe remains neutral at $z\sim7-8$ from the fraction of UV-selected galaxies that display \lya\ emission \citep{mason18,jung20,morales21}. Measurements of the cosmic microwave background suggest a midpoint at $z \simeq 8$ \citep{planck2020}.  The thermal history of the IGM down to $z \sim 5$ also suggests that much of reionization occurred at $z \sim 7$--8. \citep{boera19,gaikwad20}


Until recently, reionization was thought to be essentially complete by $z\sim6$ due to the observed onset of \lya\ transmission in quasar spectra \citep{fan06}.  On the other hand, a large scatter in \lya\ opacity has been observed in quasar sightlines at $z\leq6$ \citep{fan06,becker15,bosman18,eilers18,yang20,bosman21,zhu21,zhu22}.  The \lya\ forest at these redshifts exhibits highly opaque \lya\ and \lyb\ ``troughs'' down to $z \simeq 5.3$, the most extreme example of which is a 110 $h^{-1}$ Mpc \lya\ trough observed towards ULAS J0148+0600 \citep{becker15}.  Both these troughs and the overall scatter in \lya\ opacity have been shown to be inconsistent with a fully reionized IGM in which the ultraviolet background (UVB) is homogeneous (\citealt{becker15,bosman18,eilers18,yang20,bosman21,zhu21,zhu22}, see also \citealt{lidz06}).

The scatter in \lya\ opacity and the presence of highly opaque sightlines such as that towards ULAS J0148+0600 suggests that there are large-scale variations in the hydrogen neutral fraction at these redshifts. For an ionized IGM, the neutral hydrogen fraction is set by the photoionization rate, the gas temperature, and the total hydrogen density, which broadly suggests multiple scenarios. The first is that large-scale fluctuations in the UVB are the primary cause of the scatter in \lya\ opacity \citep{davies16,nasir20}. In this scenario, we would qualitatively expect a transmissive sightline to span a high-density region, in close proximity to ionizing sources.  In contrast, opaque sightlines would more typically be associated with voids. The second is that the scatter in \lya\ opacity is primarily driven by large-scale fluctuations in temperature \citep{daloisio15}. In this scenario, a transmissive region would be underdense, recently reionized, and hot, whereas an opaque region would have been reionized early due to its high density of ionizing sources and able to cool for longer, producing a higher recombination rate. Lastly, it is possible that reionization is still ongoing at $z<6$ and highly opaque troughs like that towards ULAS J0148+0600 correspond to islands of neutral hydrogen that have not yet been reionized \citep{kulkarni19,keating20a,nasir20}. This ``ultra-late'' reionization scenario is not mutually exclusive with the other factors; fluctuations in the UVB and temperature would still be expected.

There are a number of models that make use of these physical processes to explain the observed scatter in \lya\ opacity. Notably, the predictions they make for the relationship between opacity and density can be tested with observations. Fluctuating UVB models have been considered by numerous authors, and there are galaxy-driven variations \citep{davies16,nasir20}
and quasar-driven variations \citep{chardin15,chardin17}. We note that because quasars are rare, in quasar-driven UVB models the \lya\ opacity is less tightly coupled to density than it is in galaxy-driven UVB models. The quasar-driven model is independently disfavored because the observed number density of quasars is not high enough to produce the required number of ionizing photons for quasars to be the main sources driving reionization \citep{mcgreer18, kulkarni19,faisst22}. Additionally, a quasar-driven hydrogen reionization may be incompatible with current constraints on helium reionization \citep{daloisio17,mcgreer18,garaldi19}. Similarly, the temperature model \citep{daloisio15} is disfavored, at least as an explanation for the full range of opacities, by the observations of  \citet{becker18}, \citet{kashino20}, \citet{christenson21}, and \citet{ishimoto22}, who found that highly opaque quasar sightlines are associated with galaxy underdensities. The late reionization models commonly include UVB fluctuations, but are distinct from pure UVB models in that regions of the IGM are still significantly neutral below $z = 6$.  In these models, some highly opaque quasar sightlines correspond to neutral islands \citep{keating20b}. 
On the other hand, \citet{nasir20} find that transmissive sightlines span a range of galaxy densities, but tend towards higher values. However, $\sim 10-15$\% of transmissive sightlines in those models correspond to galaxy underdensities. \citet{keating20b} argue that sightlines where high transmission is correlated with galaxy underdensity should correspond to regions that are hot and recently reionized. 

Observations spanning a range of \lya\ opacity is necessary to robustly test the predictions from these reionization models and characterize the $z\sim 5.7$ opacity-density relationship. Previous studies have linked highly opaque quasar sightlines to galaxy underdensities towards the quasars ULAS J0148+0600 \citep{becker18,kashino20,christenson21},SDSS J1250+3130 \citep{christenson21}, and SDSS J1630+4012 \citep{ishimoto22}. \citet{ishimoto22} also observe two sightlines of lower opacity, SDSS J1137+3549 and SDSS J1602+4228, and find that they correspond to galaxy overdensities. 

In this paper, we extend our observations to some of the most highly transmissive sightlines known at these redshifts.  We characterize the density of Lyman-$\alpha$ emitting galaxies (LAEs) towards the quasars SDSS J1306+0356, which has a \lya\ effective opacity of $\tau_{\rm eff}=2.6$, and PSO J359-06, which has a \lya\ effective opacity of $\tau_{\rm eff}=2.7$, both measured over 50 $h^{-1}$ Mpc windows centered at $z=5.7$, the redshift at which we select LAEs. We additionally include new selections of LAEs in the J0148 and J1250 fields, previously published in \citet{becker18, christenson21} to make comparisons between the four fields as self-consistent as possible. We summarize the observations in Section~\ref{sec:obs}, and describe the photometry and LAE selection criteria in Section~\ref{sec:methods}.  We present the results of LAE selections in Section~\ref{sec:results}, and compare the results to predictions from current models in Section~\ref{sec:analysis} before summarizing in Section~\ref{sec:summary}. Throughout this work, we assume a $\Lambda$CDM cosmology with $\Omega_m=0.3$, $\Omega_{\Lambda}=0.7$, and $\Omega_b=0.048$. All distances are given in comoving units, and all magnitudes are in the AB system.

\section{Observations}\label{sec:obs}
\subsection{QSO Spectra}\label{sec:spectra}

\LongTables
\begin{deluxetable}{lcccc}
\tablewidth{0.5\textwidth}
\tablecaption{Effective opacity measurements for QSO sightlines referenced in this work}\label{tab:taus_summary}
\tabletypesize{\scriptsize}
\tablehead{QSO  & $z_{QSO}$ & $\tau_{\rm eff}^{50,a}$ & $\tau_{\rm eff}^{28,b}$}
\startdata
ULAS J0148+0600 & 5.998 & $7.573^c$ & $7.329^c$\\ 
SDSS J1250+3130  & 6.137 &$5.876^c$ & $5.610^c$\\
SDSS J1306+0356  & 6.0330 &$2.662\pm0.009$ & $2.475\pm0.010$ \\
PSO J359-06  & 6.1718 &$2.680\pm0.009$ & $2.392\pm0.009$\\
SDSS J1602+4228$^d$  &6.079 & $3.063\pm0.038$ & $4.898\pm0.308$\\
SDSS J1137+3549$^d$  &6.007 & $2.904\pm0.040$ & $4.344\pm0.227$\\
SDSS J1630+4012$^d$ & 6.055$^e$ & $3.857\pm0.184$ & $4.550\pm0.477$
\enddata
\tablenotetext{a}{Effective opacity measured over a 50 $h^{-1}$ Mpc window centered at 8177 \AA}
\tablenotetext{b}{Effective opacity measured over the FWHM of the $NB816$ filter (a 28 $h^{-1}$ Mpc window) centered at 8177 \AA}
\tablenotetext{c}{Lower limit}
\tablenotetext{d}{From \citet{ishimoto22}; see Section \ref{sec:sd-tau} for a detailed discussion of these sightlines}
\tablenotetext{e}{Redshift measurement from \citet{becker19}.}

\end{deluxetable}

~\label{tab:taus_summary}

\begin{figure*}
\includegraphics[width=\textwidth]{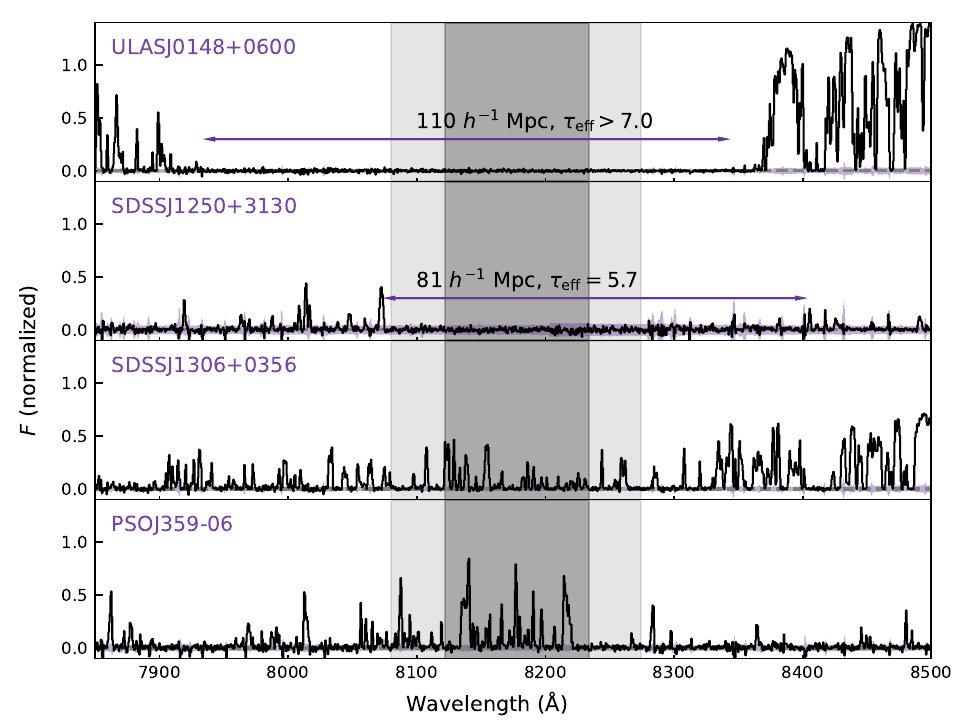}
\caption{Partial spectra of the \lya\  forest of quasars ULAS J0148+0600 (X-Shooter), SDSS J1250+3130 (Keck/ESI), SDSS J1306+0356 (X-Shooter), and PSO J359-06 (X-Shooter), whose fields we observe with Subaru/HSC. The J0148 and J1250 sightlines have $\tau_{\rm eff} \geq 7.0$ and $\tau_{\rm eff} = 5.7 \pm 0.4$ measured over 110 and 81 $h^{-1}$ Mpc respectively (trough extent marked with purple arrows). The shaded purple regions indicate the $\pm1\sigma$ uncertainty interval. The darker shaded gray rectangles indicates the FWHM of the NB816 filter, and the lighter shaded regions indicate a 50 $h^{-1}$ Mpc interval, both centered at 8177 \AA; these windows are used to calculate the effective opacity of the sightlines. The effective opacity measurements are summarized in Table 1. These spectra are normalized using PCA fits to their continuum. Note that for the J0148 spectrum, flux at $\lambda > 8350$ \AA\ is part of the quasar proximity zone and not fully normalized.}
\label{fig:qso_spectrum}
\end{figure*}

The four sightlines whose fields we survey in this work were drawn from the sample of \citet{zhu21}. This sample includes spectra of 55 quasars over $5.5\leq z \leq 6.5$ taken with the X-Shooter spectrograph on the Very Large Telescope and the Echellete Spectrograph and Imager on Keck, 23 of which are from the XQR-30 VLT Large Programme. Subsets of the four quasar spectra are shown in Figure \ref{fig:qso_spectrum}, displaying the highly opaque troughs (J0148 and J1250) and the highly transmissive regions (J1306 and J359) near $z=5.7$  found in these sightlines \citep{becker15,becker19,zhu21}. The J1306 sightline has an effective opacity of $\tau_{\rm eff}^{50}= 2.617 \pm 0.009$, where $\teff=-\rm{ln}\langle$ $F$ $\rangle$ and $F$ is the mean continuum-normalized flux. The J359 sightline has an effective opacity of $\tau_{\rm eff}^{50}= 2.661 \pm 0.009$. For both sightlines, $\tau_{\rm eff}^{50}$ is measured over 50 $h^{-1}$ Mpc windows centered at $z=5.7$ (8177 \AA), which cover $5.632<z<5.794$. These two sightlines are some of the most highly transmissive sightlines known at these redshifts \citep{zhu21}. Similarly, the J0148 and J1250 are two of the most highly opaque sightlines observed at these redshifts, with large troughs of $\tau_{\rm eff} \geq 7$ measured over 110 $h^{-1}$ Mpc and $\tau_{\rm eff} \geq 5.7 \pm 0.4$ measured over 81 $h^{-1}$ Mpc respectively. Over 50 $h^{-1}$ Mpc windows centered at 8177\AA, the sightlines have $\tau_{\rm eff}^{50} \geq 7.0$ and $\tau_{\rm eff}^{50}=5.03 \pm 0.21$ respectively.  We also calculate $\tau_{\rm eff}$ for these sightlines over a 28 $h^{-1}$ Mpc window, which represents the full width at half maximum of the narrowband filter used for LAE selection, and find $\tau_{\rm eff}^{28} \geq 7.329$ for the J0148 sightline, $\tau_{\rm eff} \geq 5.610$ for the J1250 sightline, $\tau_{\rm eff} = 2.475\pm0.010$ for the J1306 sightline, and $\tau_{\rm eff} = 2.392\pm 0.009$ for the J359 sightline. For all of the effective opacity measurements, we adopt a lower limit of $\tau_{\rm eff} \geq -\ln(2 \sigma_{ \langle F \rangle})$ if the mean flux is negative or detected with less than $2\sigma$ significance. This definition is consistent with previous works (e.g., \citealt{eilers18}). The opacity measurements used in this work are summarized in Table 1. We estimate \lya\ opacity for these sightlines using our imaging data in Appendix A.



\subsection{HSC Imaging}\label{sec:imaging}

Presented here for the first time are imaging data in the J1306 and J359 fields, taken with Subaru Hyper Suprime Cam (HSC). This work also makes use of HSC imaging in the J0148 and J1250 fields, previously presented in \citet{becker18} and \citet{christenson21}. Observations of the J1306 field were made via the HSC queue in April and June 2019, May 2020, and January and June 2021. Observations of the J359 field were made via the HSC queue in October and November 2019, August 2020, and November 2021. All observations in these fields were made during dark time. As for previous fields surveyed in this program, images were centered at the quasar position. This program makes use of two HSC broadband filters, $r2$ and $i2$, and the narrowband $NB816$ filter, which has a transmission-averaged mean wavelength $\lambda=8168$ \AA\ and $\geq50$\% transmission over $8122$ \AA $\leq \lambda \leq$ 8239 \AA. The wavelength coverage of the $NB816$ filter coincides with the \lya\ line at $z\sim5.7$.

\begin{center}
\begin{deluxetable}{lcccccc}[ht]
\tablewidth{1.75\textwidth}
\tablecaption{Summary of HSC imaging \label{tab:obs_summary}}
\tablehead{\colhead{} & \colhead{Filter} & \colhead{$t_{exp}$ (hrs)} &
\colhead{ Seeing$^a$} & \colhead{$m_{50\% >5\sigma}^b$} & $m_{5\sigma, 1.2\arcsec}^c$ }
\startdata
\multirow{3}{*}{J0148}&$r2$ & 1.5 & 0.61 & 26.3 & 26.2 \\*
&$i2$ & 2.4 & 0.71 & 25.9 & 25.8\\*
&$NB816$ & 4.5 & 0.60 & 25.1 & 25.2 \\*
\hline
\multirow{3}{*}{J1250}&$r2$ & 2.0$^d$ & 1.07 & 26.3 & 26.2 \\*
&$i2$ & 2.5 & 0.62 & 26.1 & 26.0 \\*
&$NB816$ & 2.8 & 0.73 & 25.1 & 25.2\\*
\hline
\multirow{3}{*}{J1306}&$r2$ & 1.3 & 0.89 & 26.3 & 26.2\\*
&$i2$ & 2.4 & 0.74  & 26.1 & 26.0\\*
&$NB816$ & 2.8 & 0.80  & 25.0 & 25.1 \\*
\hline
\multirow{3}{*}{J359}&$r2$ &1.5 & 1.08 & 26.2 & 26.3\\*
&$i2$ & 1.9 & 0.73  & 25.8 & 25.9\\*
&$NB816$ & 2.2 & 0.87  & 25.0 & 25.2
\enddata
\tablenotetext{a}{Median seeing FWHM in combined mosaic.}
\tablenotetext{b}{Magnitude at which 50\% of detected sources have $S/N \geq 5$ in the corresponding filter.}
\tablenotetext{c}{Limiting magnitude, given by five times the standard deviation of the flux measured in empty 1.5\arcsec\ apertures.}
\tablenotetext{d}{Partially observed during gray time.}

\end{deluxetable}
\end{center}

~\label{obs_summary}

The observations in all four fields are summarized in Table \ref{tab:obs_summary}, as well as the image depth measured in empty 1.2\arcsec\ apertures and the median $5\sigma$ limiting aperture magnitudes in each band. At the limiting magnitudes, at least 50\% of the detected sources have a signal-to-noise ratio $S/N_{NB816} \geq 5$.

We used version 21 of the LSST Science Pipeline \citep{ivezic08,juric15} to reduce individual CCDs and combine them into stacked mosaics. The pipeline uses PanStarrs DR1 imaging \citep{chambers16} for photometric calibrations. We used Source Extractor \citep{bertin96} to create a catalog of NB816-detected sources and their spatial coordinates in the stacked mosaics, and then make our own photometric measurements at these coordinates, as described below.

\section{Methods}\label{sec:methods}

\subsection{Photometry}\label{sec:photometry}
The LAE selection in this work makes use of aperture fluxes as the primary photometric measurement. This choice is a departure from our previous work, which was based on PSF  \citep{christenson21} or CModel fluxes \citep{becker18}. As we discuss further in Section \ref{sec:analysis}, a major focus of this paper is comparing the fields to one another, which requires minimizing the effect of variations in depth, seeing, and completeness. While PSF fluxes can be optimized for the detection of faint and unresolved sources, aperture fluxes are less easily impacted by small changes in the seeing and more robust for resolved sources. We have therefore opted to accept a lower signal-to-noise ratio and the loss of some faint LAEs from our catalog in favor of a more robust selection.

The source detection and photometric measurements are carried out via the following steps. Source positions are identified in the NB816 stacked mosaic using Source Extractor. At each source position, we measure the flux in a 1.5\arcsec\ aperture, and also measure the sky background in a $1.5-5$\arcsec\ annulus around the aperture, excluding pixels labeled as sources by the LSST pipeline. The aperture fluxes are corrected by the measured sky background. These measurements are made independently, at the same position, in each band.

\subsection{LAE selection procedure}\label{sec:lae_selection}

Our selection criteria, following \citet{christenson21}, are based on those of \citet{ouchi08}. As noted in \citet{christenson21}, our observations have some disparity in depth between different bands and fields. To ensure a high-quality selection of LAEs, we impose additional requirements that are designed to exclude objects with large uncertainties in their colors.  The selection criteria are as follows:

\begin{itemize}
\item $NB \leq 25.5$
\item $S/N_{NB816} \geq 5.0$
\item $\frac{F_{NB816}}{F_{i2}} \geq 3.0$ (50\% probability) and $\frac{F_{NB816}}{F_{i2}} \geq 1.7$ (95\% probability)
\item $F_{r2} \leq 2\sigma_{r2}$, or $F_{r2}\geq2\sigma_{r2}$ and $F_{i2}/F_{r2}\geq 2.5$ 
\item $\frac{F_{NB816}}{F_{r2}} \geq 7.6$ (50\% probability) and $\frac{F_{NB816}}{F_{r2}} \geq 4.0$ (95\% probability)
\end{itemize}

\begin{figure*}
\includegraphics[width=\textwidth]{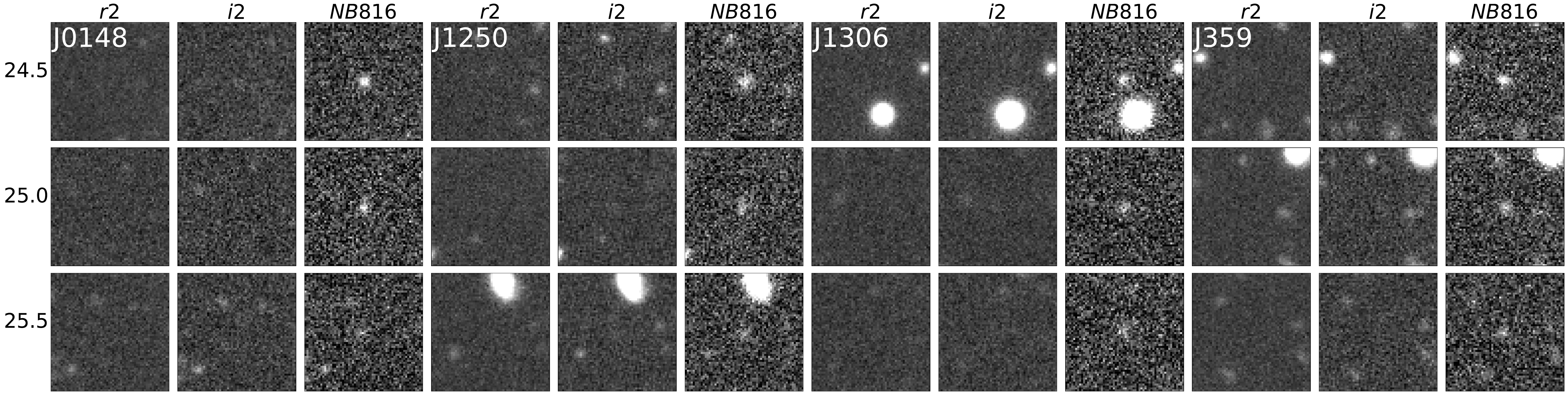}
\caption{Example LAE candidates selected in all four fields with the criteria described in Section \ref{sec:lae_selection}. The cutout images are 10 \arcsec\ on each side and centered on the LAE position. For each field, we show an example candidate selected to have NB816 = 24.5, 25.0, and 25.5 (top to bottom) in the $r2, i2$, and NB816 bands (left to right).}
\label{fig:lae_stamps}
\end{figure*}

The 95\% probability thresholds are the lower bound of the $1\sigma$ error of an object with $F_{NB816}/F_{i2} = 3.0$, $F_{i2}/F_{r2}=2.5$, and $S/N_{NB816} = 5.0$. All objects that satisfy these requirements undergo a visual inspection to remove spurious sources, such as satellite trails. Examples of selected LAEs are shown in Figure \ref{fig:lae_stamps}.

\subsection{Completeness Corrections}\label{sec:completion}
\begin{figure*}
\includegraphics[width=\textwidth]{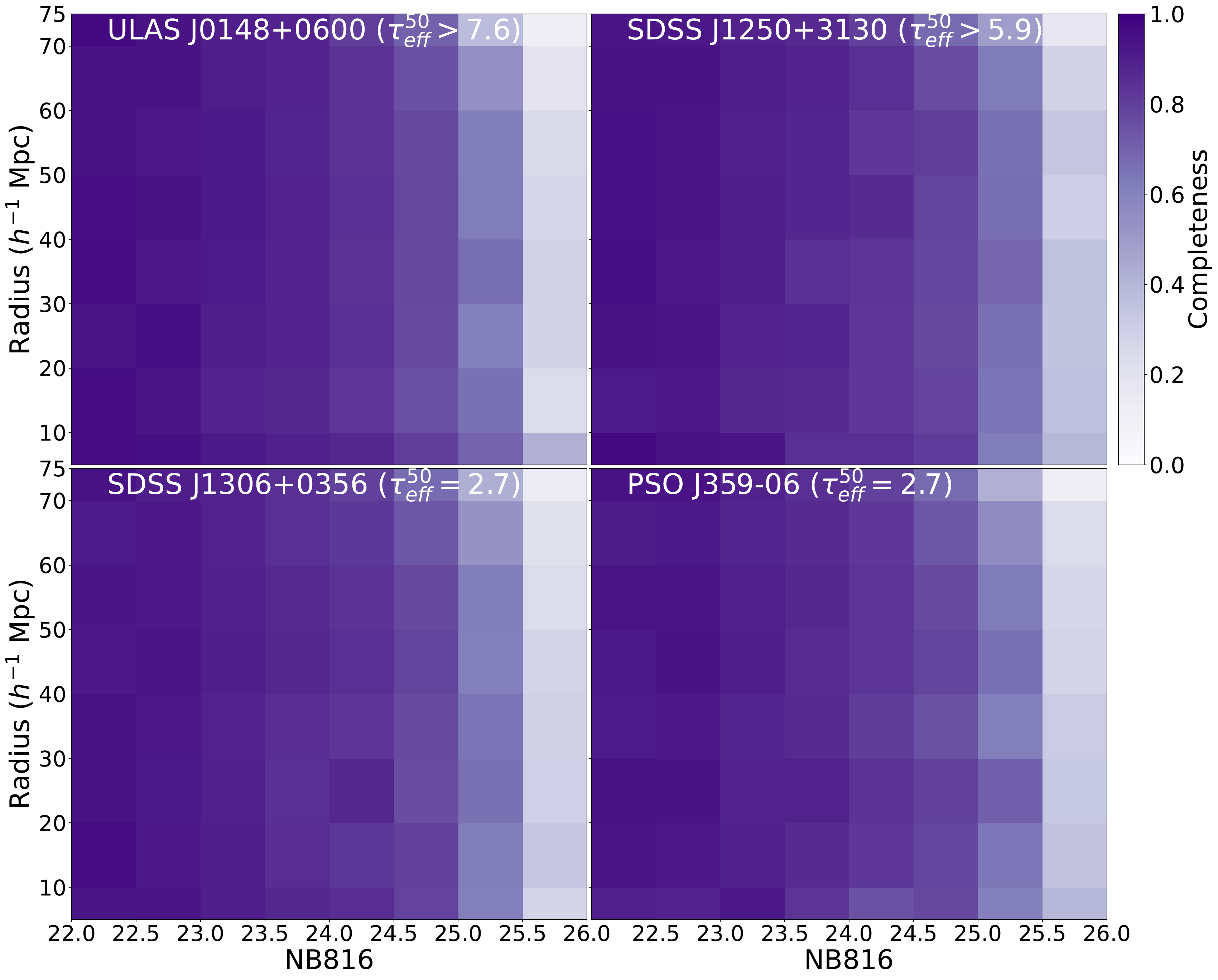}
\caption{Completeness measured in the J0148 (top left), J1250 (top right), J1306 (bottom left) and J359 (bottom right) fields as a function of projected distance from the quasar position and $NB816$ magnitude. The completeness is given by the fraction of artificial LAEs injected into the imaging that were detected by our LAE selection procedure. Note that we have included narrowband magnitudes down to $NB816=26.0$; however, we only include sources down to $NB816=25.5$ due to the low completeness in the faintest magnitude bin.}
\label{fig:completeness}
\end{figure*}

\begin{figure}
\includegraphics[width=0.49\textwidth]{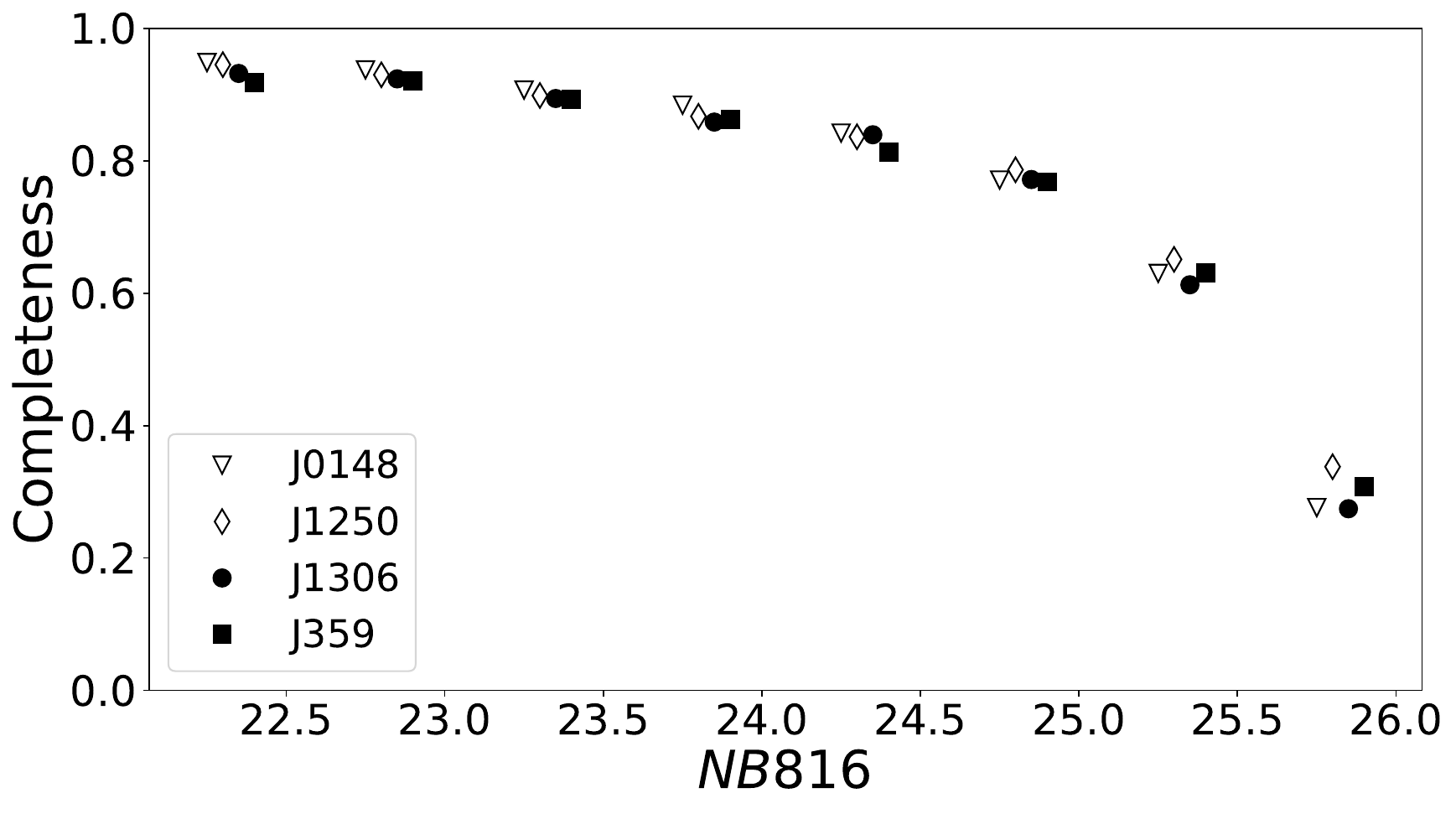}
\caption{Radially averaged completeness measured in the J0148 (filled gray triangles), J1250 (open gray diamonds), J1306 (filled black circles), and J359 (open black squares) fields as a function of $NB816$ radius as a function of distance from the quasar position and $NB816$ magnitude. The fields are offset horizontally for clarity. While we have calculated the completeness for narrowband magnitudes down to $NB816=26.0$, our analysis only includes sources down to $NB816=25.5$.}
\label{fig:completeness_mag}
\end{figure}

We make completeness corrections for the selected catalog of LAEs in two stages. The first stage is to calculate a completeness correction in each field as a function of $NB816$ magnitude and distance from the quasar sightline. This calculation is based on artificially injected LAE candidates, which are placed at randomly generated positions in the field, binned by radius and magnitude, put through the LAE selection procedures described in Section \ref{sec:methods}. The sample is generated such that the artificial LAEs are spread roughly evenly between the magnitude and radius bins, ensuring that there are enough objects in each bin to calculate a completeness correction. The completeness correction is the reciprocal of the fraction of artificial LAEs that were successfully detected in each bin. We show the completeness measured in each field as a function of distance from the quasar position and NB816 magnitude in Figure \ref{fig:completeness}. The completeness calculations are made down to NB816 $\leq 26.0$, but we only select LAEs to NB816 $\leq 25.5$ in our final catalog because of the low completeness measured in the faintest magnitude bin. This completeness correction is used to correct the measured surface density as a function of radius and magnitude shown in Section \ref{sec:results} in Figures \ref{fig:sd_mag}, \ref{fig:sd_rad}, \ref{fig:sd_tau_field}, and \ref{fig:sd_tau_allfields}. We additionally show the radially averaged completeness as a function of $NB816$ magnitude in Figure \ref{fig:completeness_mag}.

The second stage is calculating total completeness as a function of position in each field. We use the completeness-corrected magnitude distribution of LAEs detected in all four fields to generate a second set of artificial LAEs in each field, this time with NB816 magnitudes drawn from the empirical magnitude distribution. Because these artificial LAEs are representative of the real LAE sample, we can use them to calculate a map of completeness as a function of position. We assign each artificial LAE a flag indicating whether or not it was successfully selected using our LAE selection procedure, and then calculate the surface density of both (i) the full artificial LAE catalog and (ii) the selected artificial LAEs in each field as a function of position. Surface densities are estimated using the kernel density estimation approach described below. The completeness as a function of position is then given by the surface density of the selected LAEs divided by the surface density of the injected LAEs. The completeness is fairly uniform, with variations of $\leq10$\% on $\sim$ 5 $h^{-1}$ Mpc scales, out to a radius of $\sim$40\arcmin. At larger radii, the completeness declines sharply. We calculate these completeness correction maps separately for each field, and apply them to the LAE maps shown in Figure \ref{fig:lae_map}.

\section{Results}\label{sec:results}
\begin{figure}
\includegraphics[width=0.49\textwidth]{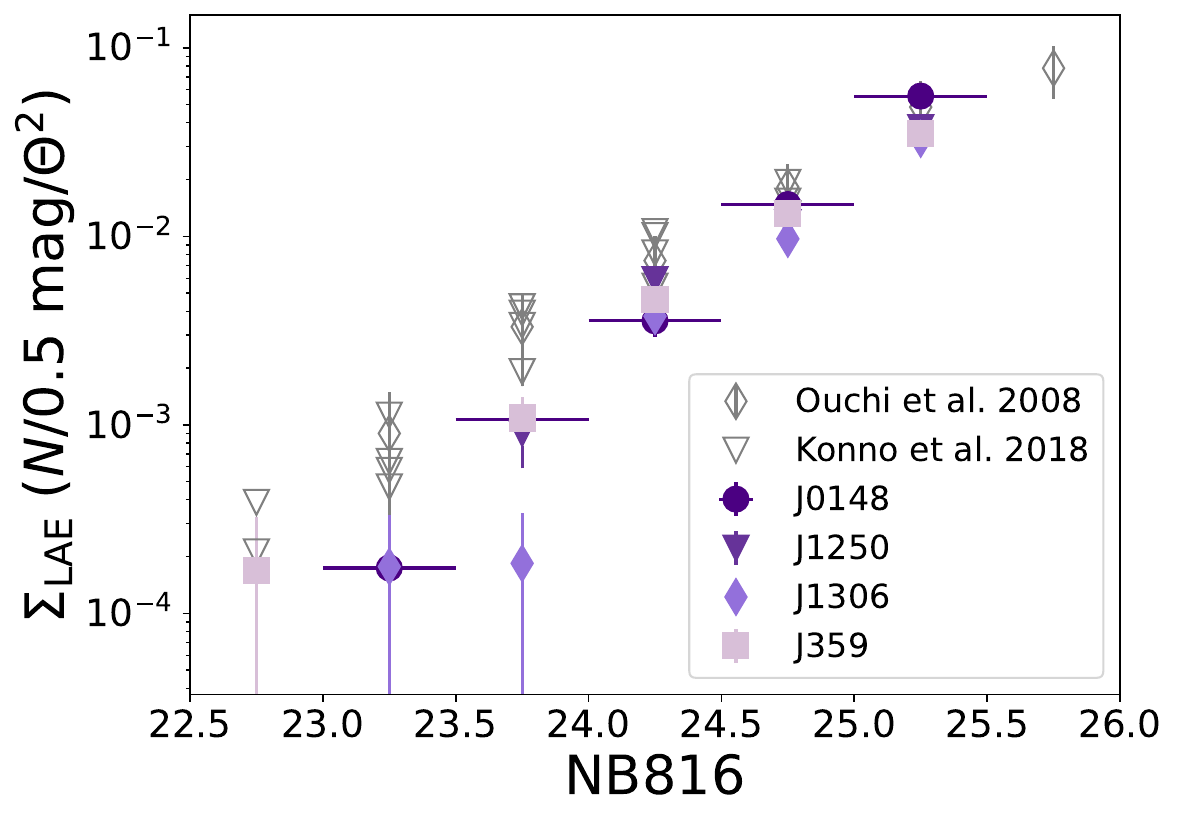}
\caption{Completeness-corrected surface density of LAE candidates in the J0148, J1250, J1306, and J359 fields (filled markers) as a function of their $NB816$ magnitude (see Section \ref{sec:completion} for details on the completeness correction.). The error bars on the completeness-corrected measurements are 68\% Poisson intervals. We also show measurements from \citet{konno18} (open gray triangles, includes four HSC fields plotted separately) and \citet{ouchi08} (open gray diamonds) for comparison.}
\label{fig:sd_mag}
\end{figure}

We select 298 LAEs in the J0148 field, 247 in the J1250 field, 192 in the J1306 field, and 228 in the J359 field using the procedures outlined in section \ref{sec:methods}. The number of LAEs selected in the J0148 and J1250 fields is somewhat lower than found in \citet{christenson21} due to the use of aperture fluxes, although the spatial distribution of the sources is qualitatively similar. We compare the two selections in more detail in Appendix \ref{sec:c21}. We show the completeness-corrected surface density of the LAE catalogs in each field as a function of their NB816 magnitude in Figure \ref{fig:sd_mag}.  Also included are measurements from \citet{konno18} and \citet{ouchi08}. We note that we find a lower surface density of bright objects than reported in the literature. This difference arises from our use of 1.5\arcsec\ apertures throughout, rather than using adaptively scaled apertures for the primary magnitude measurement as done in \citet{ouchi08}.  In some cases the 1.5\arcsec\ apertures miss some of the flux in brighter, more extended objects.

\begin{center}
\begin{deluxetable}{llccc}[!]
\tablewidth{2\textwidth}
\tablecaption{LAE number density as a function of radius \label{tab:sd_rad}}
\tablehead{\colhead{} & \colhead{R (Mpc)}  & \colhead{$N_{LAEs}$} & \colhead{$N_{corr}^a$} & \colhead{$\Sigma$ LAE (Mpc h$^{-1}$)$^{2,b}$} }
\startdata
\multirow{8}{*}{J0148} & $5 (0-10)$  &0 & 0 & 0.0  ( 0.0 $-$ 0.0 )\\
&$15 (10-20)$  & 12 & 17 & 0.018  ( 0.013 $-$ 0.022 )\\
&$25 (20-30)$  & 33 & 50 & 0.032  ( 0.028 $-$ 0.036 )\\
&$35 (30-40)$  &  44 & 61 & 0.028 ( 0.024 $-$ 0.031 )\\
&$45 (40-50)$  &  51 & 76 & 0.027  ( 0.024 $-$ 0.03 )\\
&$55 (50-60)$  & 56 & 86 & 0.025  ( 0.022 $-$ 0.027 )\\
&$65 (60-70)$  &  74 & 126 & 0.031  ( 0.028 $-$ 0.034 )\\
&$72 (70-74.5)$  &  28 & 61 & 0.03  ( 0.026 $-$ 0.034 )\\

\hline
\multirow{8}{*}{J1250}&$5 (0-10)$  & 2 & 3 & 0.01  ( 0.005 $-$ 0.016 )\\
&$15 (10-20)$  & 9 & 12 & 0.013  ( 0.009 $-$ 0.017 )\\
&$25 (20-30)$  & 21 & 29 & 0.019  ( 0.015 $-$ 0.022 )\\
&$35 (30-40)$  & 32 & 44 & 0.02  ( 0.017 $-$ 0.023 )\\
&$45 (40-50)$  & 37 & 52 & 0.018  ( 0.016 $-$ 0.021 ) \\
&$55 (50-60)$  & 61 & 87 & 0.025  ( 0.022 $-$ 0.028 )\\
&$65 (60-70)$  & 63 & 94 & 0.023  ( 0.021 $-$ 0.025 ) \\
&$72 (70-74.5)$  &22 & 37 & 0.018  ( 0.015 $-$ 0.021 )\\
\hline
\multirow{8}{*}{J1306}&$5 (0-10)$  & 2 & 3 & 0.01  ( 0.005 $-$ 0.016 ) \\
&$15 (10-20)$  & 16 & 24 & 0.025  ( 0.02 $-$ 0.031 )\\
&$25 (20-30)$  & 28 & 40 & 0.026  ( 0.022 $-$ 0.029 )\\
&$35 (30-40)$  & 35 & 50 & 0.023  ( 0.02 $-$ 0.026 ) \\
&$45 (40-50)$  & 27 & 40 & 0.014  ( 0.012 $-$ 0.016 )\\
&$55 (50-60)$  & 39 & 59 & 0.017  ( 0.015 $-$ 0.019 )\\
&$65 (60-70)$  & 30 & 50 & 0.012  ( 0.01 $-$ 0.014 ) \\
&$72 (70-74.5)$  & 15 & 31 & 0.015  ( 0.013 $-$ 0.018 )\\ 
\hline
\multirow{8}{*}{J359}&$5 (0-10)$  & 1 & 1 & 0.004  ( 0.002 $-$ 0.006 ) \\
&$15 (10-20)$  & 11 & 16 & 0.017  ( 0.012 $-$ 0.021 )\\
&$25 (20-30)$  & 22 & 29 & 0.019  ( 0.015 $-$ 0.022 )\\
&$35 (30-40)$  & 32 & 49 & 0.022  ( 0.019 $-$ 0.025 ) \\
&$45 (40-50)$  & 34 & 48 & 0.017  ( 0.015 $-$ 0.019 )\\
&$55 (50-60)$  & 46 & 69 & 0.02  ( 0.018 $-$ 0.023 )\\
&$65 (60-70)$  & 57 & 89 & 0.022  ( 0.02 $-$ 0.024 ) \\
&$72 (70-74.5)$  & 25 & 43 & 0.021  ( 0.018 $-$ 0.024 )\\
\enddata
\tablenotetext{a}{Completeness corrected}
\tablenotetext{b}{Ranges quoted in parentheses correspond to 68\% Poisson intervals.}
\end{deluxetable}
\end{center}

\label{sd_rad}

\begin{figure*}
\includegraphics[width=\textwidth]{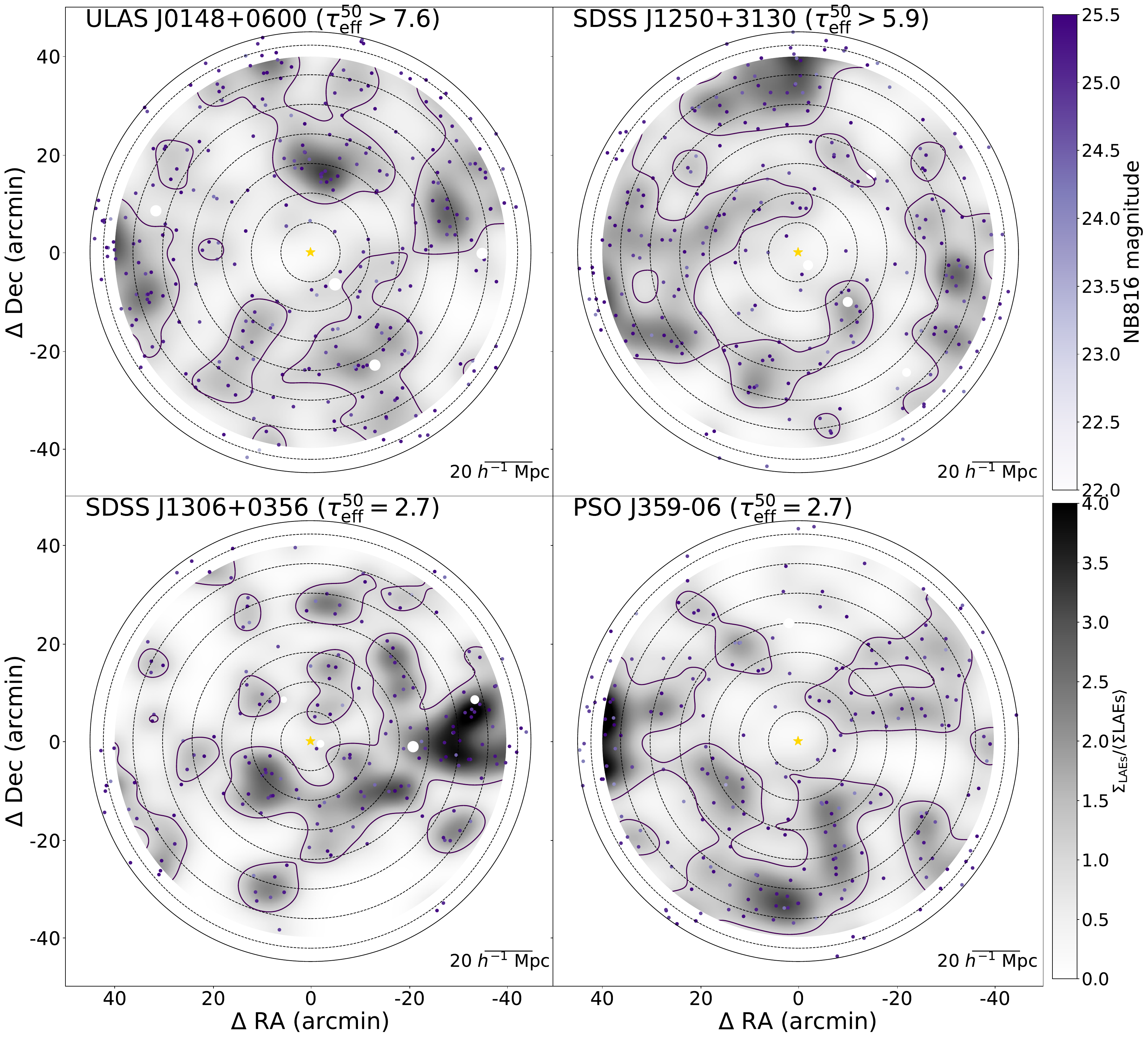}
\caption{Distribution of LAE candidates in all four fields: J0148 (top left), J1250 (top right), J1306 (bottom left), and J359 (bottom right). The LAE candidates are assigned a color that indicates their NB816 magnitude. The grayscale shading in the background indicates the surface density of LAE candidates, which we calculate by kernel density estimation and normalized by the mean surface density of each field, measured over $15\arcmin \leq \Delta \theta \leq 40\arcmin$. This surface density is corrected for spatial variations in completeness as described in Section \ref{sec:completion}. The field is centered on the quasar position, which is marked with a gold star, and the concentric dotted rings indicate 10 $h^{-1}$ Mpc intervals from the quasar position. The solid ring marks the edge of the field, 45\arcmin\ from the quasar position. Portions masked out of the field in white are obscured by foreground stars.}
\label{fig:lae_map}
\end{figure*}

Figure \ref{fig:lae_map} shows the distribution of LAE candidates in all four fields: J0148 (top left), J1250 (top right), J1306 (bottom left), and J359 (bottom right). In each panel, the field is centered on the quasar position, which is marked with a yellow star. The concentric dotted rings indicate 10 $h^{-1}$ Mpc intervals from the quasar position, and the solid black ring indicates the edge of the field. The LAEs are represented with a color that indicates their NB816 magnitude.  There are several bright foreground stars in these fields that obscure small portions of the field, which are masked out in white. The grayscale shading indicates the surface density of LAEs. To calculate the surface density, we overlay a grid of 0.24\arcmin\ (0.4 $h^{-1}$ Mpc) pixels on the field and then find the surface density in each grid cell by kernel density estimation using a Gaussian kernel with a 1.6\arcmin\ bandwidth. This smoothing scale is chosen to match the mean separation between each LAE and its nearest neighbor. The surface density is then completeness-corrected as described in Section \ref{sec:completion} and normalized by the mean surface density of the field over $15\leq \theta \leq 40$ arcmin. See Appendix C for maps normalized using a global mean surface density, calculated over the $15\leq \theta \leq 40$ arcmin region of all four fields.

\begin{figure*}
\includegraphics[width=\textwidth]{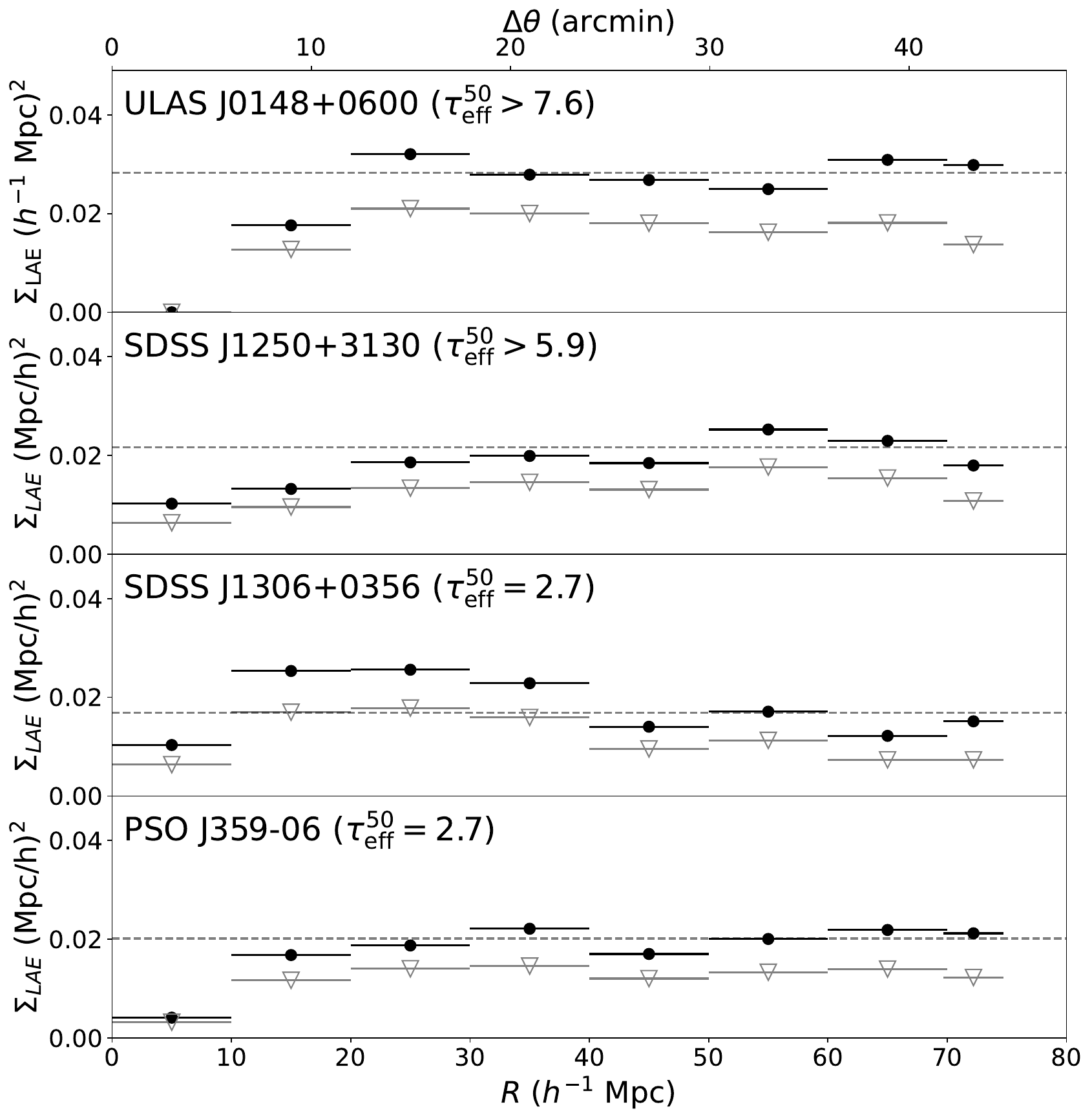}
\caption{Surface density of LAE candidates in all four fields (from top to bottom: J0148, J1250, J1306, J359) as a function of their distance from the quasar position, measured in 10 $h^{-1}$ Mpc annular bins. The unfilled gray triangles indicate raw surface density measurements, and the filled black circles indicate completeness-corrected measurements. The dotted line represents the mean completeness-corrected surface density in the field measured over $15\arcmin \leq \Delta \theta \leq 40\arcmin$. The horizontal error bars indicate the width of the annulus.}
\label{fig:sd_rad}
\end{figure*}

Figure \ref{fig:sd_rad} shows the surface density of LAEs in each field as a function of distance from the quasar position. We first measure the raw surface density by binning the LAEs into 10 $h^{-1}$ Mpc annuli, and then further bin them by NB816 magnitude to apply the completeness correction shown in Figure \ref{fig:completeness}. The raw measurements are shown in Figure \ref{fig:sd_rad} with gray, open triangles, and the completeness-corrected measurements are shown with filled, black circles. The horizontal dotted line represents the mean completeness-corrected surface density of the field, which we measure over $15\leq \theta \leq 40$\arcmin. The surface density measurements in each annular bin for the four fields are summarized in Table \ref{tab:sd_rad}. 

We find that all four fields in the survey are underdense within $\sim$ 10 $h^{-1}$ Mpc of the quasar sightline; all except the J1306 field are also underdense out to 20 $h^{-1}$ Mpc. The J1306 field is mildly overdense between 10 and $\sim30-40$ $h^{-1}$ Mpc. This re-selection of LAEs in the J0148 and J1250 fields based on aperture photometry is consistent with our previous selections in \citet{christenson21} (J0148 and J1250) and \citet{becker18} (J0148), both in the large-scale structures reflected in the LAE distribution and in the association between highly opaque sightlines and galaxy underdensities. Additionally, we newly find an association between these two transmissive sightlines and galaxy underdensities within 10 $h^{-1}$ Mpc.

\section{Analysis}\label{sec:analysis}
\subsection{Comparison of radial distributions to model predictions}\label{model_comparison}
\begin{figure*}
\includegraphics[width=\textwidth]{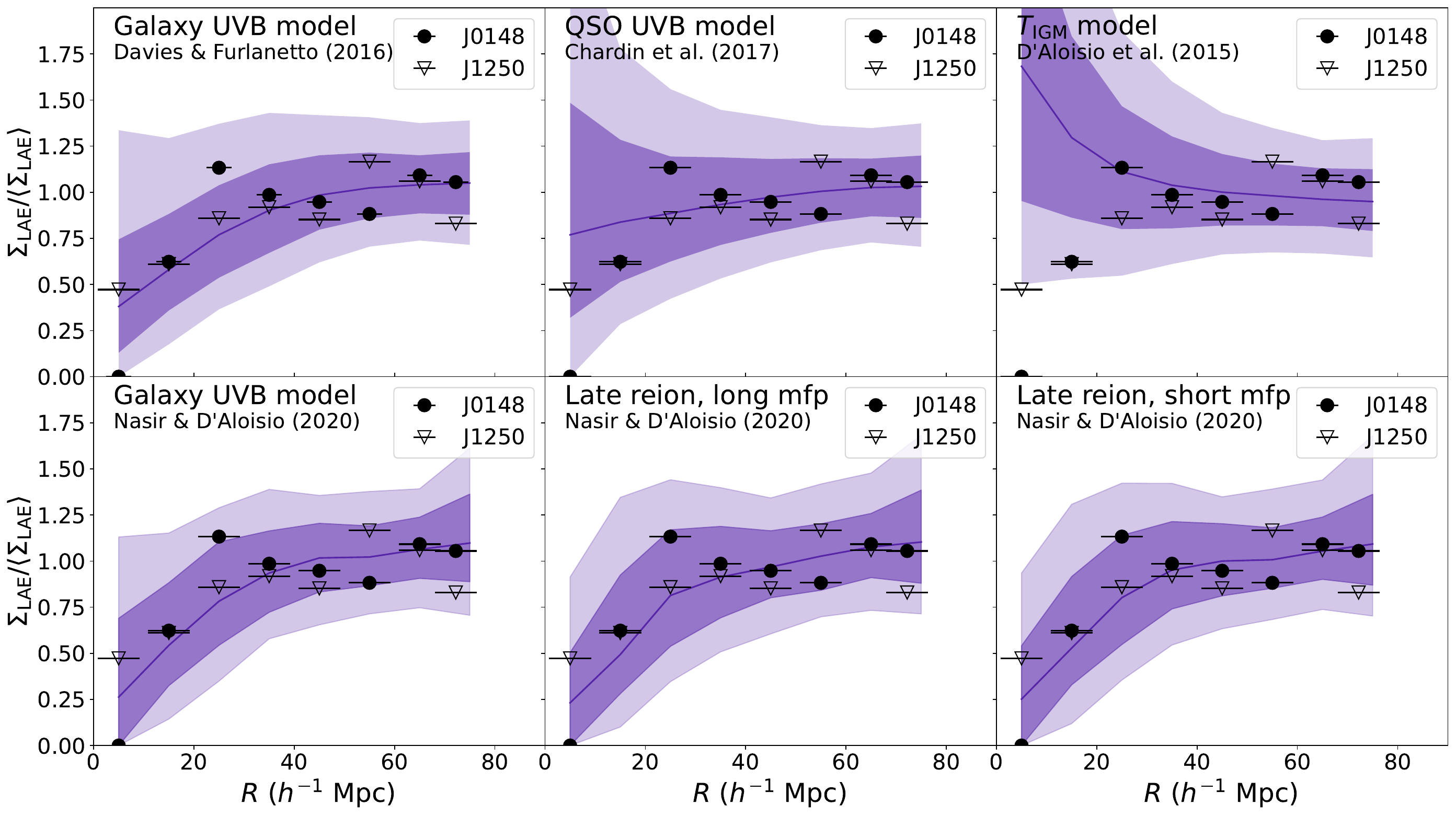}
\caption{Surface density profiles for highly opaque lines of sight.  Each panel compares the observed radial distribution of LAE candidates in the J0148 (filled circle) and J1250 (open triangle) fields to model predictions, where the model lines of sight have $\tau_{\rm eff}^{50}\geq7.0$. The top row shows predictions from the galaxy UVB model based on \citet{davies16} (top left), the QSO UVB model based on \citet{chardin15,chardin17} (top center), and the fluctuating temperature model from \citet{daloisio15} (top right).   The bottom row shows predictions from \citet{nasir20}, including their galaxy UVB (early reionization) model (bottom left), late reionization model with a long mean free path (bottom center), and late reionization model with a short mean free path (bottom right). The solid lines show the median predictions for each model. The dark- and light-shaded regions show 68\% and 98\% ranges respectively. As in Figure \ref{fig:sd_rad}, the horizontal error bars on the data points indicate the width of the bins. All surface densities are given normalized by the mean surface density in the field, measured over 15\arcmin\ $\leq\theta\leq $ 40\arcmin.}
\label{fig:models-opaque}
\end{figure*}

\begin{figure*}
\includegraphics[width=\textwidth]{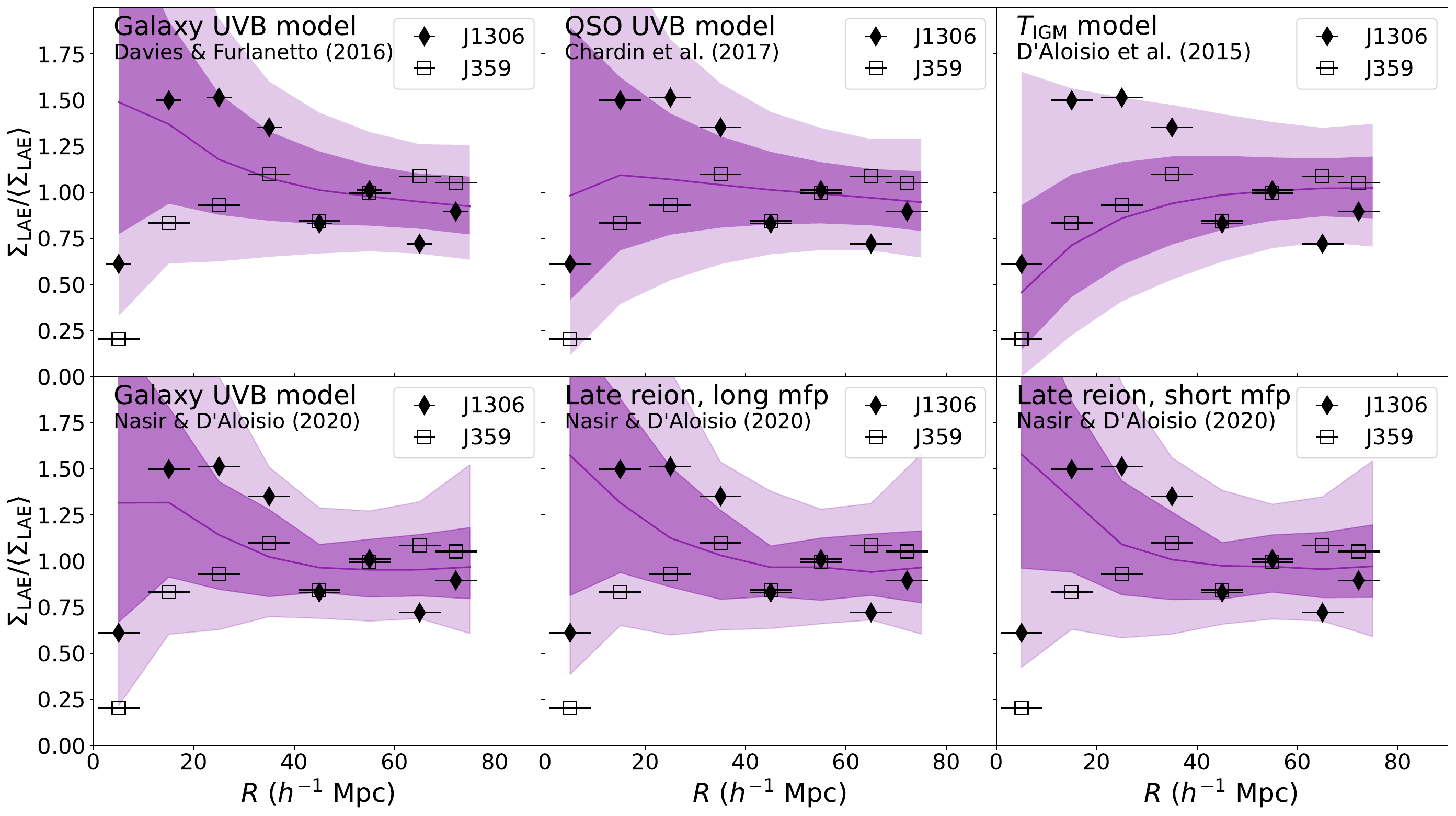}
\caption{Surface density profiles for transmissive lines of sight.  Each panel compares the observed radial distribution of LAE candidates in the J1306 (filled circle) and J359 (open triangle) fields to model predictions.
The models are the same as in Figure \ref{fig:models-opaque}, but for model lines of sight with $\tau_{\rm eff}^{50} = 2.5 \pm 0.25$. 
Lines, shading, and error bars are as in Figure \ref{fig:models-opaque}.  All surface densities are given normalized by the mean surface density in the field, measured over 15\arcmin\ $\leq\theta\leq $ 40\arcmin.}
\label{fig:models-transmissive}
\end{figure*}

We can compare the results of the LAE selection in these fields directly to the predictions made by various models. In this section, we consider only the four fields surveyed in this work. Other sightlines from the literature are discussed in Section \ref{sec:sd-tau}. 

We consider the three main types of models described in the introduction: fluctuating UVB, fluctuating temperature, and ultra-late reionization.  Of these three types of models, we consider six variations.
Two are galaxy-driven UVB models, one from \citet{davies18} and another, which also includes temperature fluctuations as would be expected at the end of reionization, from \citet{nasir20}. A third UVB model, from \citet{chardin15,chardin17}, is quasar-driven. The fourth is a fluctuating temperature model from \citet{daloisio15}. Lastly, we consider two variations on an ultra-late reionization scenario from \citet{nasir20}. These models incorporate fluctuations in temperature and UVB as expected at the end of reionization, but allow the IGM to be $\sim10$\% neutral at $z=5.5$. Of these two models, one uses a short mean free path (8 $h^{-1}$ Mpc at z=6) and the other a long mean free path (23 $h^{-1}$ Mpc at $z=6$). For comparison, \citet{becker21} measure a mean free path of 3.57 $h^{-1}$ Mpc at $z=6$. 

The predictions for surface density of LAEs as a function of radius are constructed from sightlines that have $\tau_{\rm eff}^{50}=2.5\pm 0.25$ (transmissive predictions) or $\tau_{\rm eff}^{50}\geq7.0$ (opaque predictions).  We note that the J1250 sightline has $\tau_{\rm eff}^{50} =5.033\pm0.215$, which is somewhat lower than the simulated opaque sightlines used here; however, \citet{davies18} find that model predictions for $\tau_{\rm eff}^{50} \geq 5.0$ are very similar (see also Figure \ref{fig:sd_tau_field}). For each model, simulated LAE populations around these sightlines are constructed using the following basic procedure: galaxies are assigned to dark matter halos, using the measured UV luminosity of \citet{bouwens15} for abundance matching, and their spectra are modeled as a power-law continuum with a \lya\ emission line with equivalent width set by the models of \citet{dijkstra12}. We refer the reader to \citet{nasir20} and \citet{davies18} for further details. 

To ensure that the comparison between the modelled LAE populations and our models is as close as possible, we match the surface density of the model population to that of the observed population. First, we remove simulated LAEs from the sample in a radially- and magnitude-weighted manner using the observed completeness correction to create an incomplete catalog of simulated LAEs, comparable to the raw, uncorrected observations. The completeness correction is scaled by a factor of $\sim 0.6$, so that the mean surface density of the incomplete simulated LAEs matches the uncorrected median surface density of real LAEs in our four fields. We then apply the completeness correction without the scaling factor, as done with the real LAEs, to produce a completeness-corrected simulated LAE population. From this completeness-corrected sample, we construct expected surface density profiles for highly opaque and transmissive lines of sight, which we compare to our measurements.



Figure \ref{fig:models-opaque} shows the measured surface density in the J0148 (filled circle) and J1250 (open triangle) fields as a function of radius alongside model predictions for opaque sightlines. Similarly, Figure \ref{fig:models-transmissive} shows the comparison between the measured surface density in the J1306 (filled circle) and J359 (open triangle) fields as a function of radius and model predictions for transmissive sightlines. In both sets of figures, the top row shows, from left to right, predictions from the galaxy UVB model \citep{davies18}, quasar UVB model \citep{chardin15,chardin17}, and temperature model \citep{daloisio15}. The bottom row shows, from left to right, predictions from \citet{nasir20} for the galaxy UVB model, the ultra-late reionization model with a long mean free path, and the ultra-late reionization model with a short mean free path. In each panel, the mean model prediction is shown with a solid line, and the shaded regions indicate the 68\% and 98\% ranges. All predictions and measurements measurements are normalized over the mean surface density in each field, measured over 15\arcmin\ $\leq\theta\leq$ 40\arcmin. In this figure, and in others following that compare our observations to models, we have omitted vertical error bars. The ranges we give in Table 3 are based on Poisson statistics, and including them here would imply that the field-to-field variations are based primarily on shot noise rather than cosmic variance. The purpose of these figures is to consider whether our observations could be individual realizations of these models based on the range of density values expected for individual fields in each model. The most relevant variance for this comparison is therefore the model variance.

We find that all four sightlines are underdense within 10 $h^{-1}$ Mpc of the quasar sightline, compared to the mean surface density of the exterior of the field. The highly opaque sightlines strongly disfavor the temperature model but are consistent with predictions from the UVB and late reionization models, as found in \citet{becker18} and \citet{christenson21}. There is some tension, however, between the transmissive sightlines and these models. The J359 sightline falls below the lower 98\% threshold at $R \le 10 h^{-1}$ Mpc for all four of the galaxy UVB and late reionization models, and the J1306 model falls below the lower 68\% threshold in the same ranges. This suggests that these models are unlikely to produce transmissive sightlines that are as underdense as the two we have observed. Taking all four sightlines into account, none of the models we consider here are obviously consistent with all of the data.

\subsection{Environments of extreme-opacity sightlines}\label{sec:environments}

A main focus of this paper is interpreting the four sightlines together, to consider what we can infer about the environments in which extreme opacity sightlines arise. The two highly opaque sightlines clearly show underdense regions within 20 $h^{-1}$ Mpc of the quasar sightline. Similarly, the J359 sightline sits in an underdense region that is longer, but narrower, running in roughly the east-west direction. These underdense regions have a large lateral extent, spanning tens of comoving megaparsecs. The opaque troughs extend over 160 and 80 $h^{-1}$ Mpc (J0148 and J1250 respectively), and the J359 sightline is transmissive over a 50 $h^{-1}$ Mpc segment of the \lya\ forest. We consider a region transmissive based on the absence of dark gaps ($\geq 30 h^{-1}$ Mpc in length, as defined by \citealt{zhu21}) - or, more simply, that it is populated by transmission spikes that are measurable in extent relative to the continuum level. The lateral extent of these underdensities suggests that, were they to also extend over the full lengths of the corresponding \lya\ forest features, these extreme sightlines could arise from very large structures.

The J1306 sightline arises from a region that is underdense, but adjacent to overdense regions. Approximately 45\% of the area within 20 $h^{-1}$ Mpc of the quasar sightline is estimated to be overdense (Figure \ref{fig:lae_map}), compared to 7\% (J0148), 14\% (J1250), and 19\% (J359) for the other fields. The galaxy overdensity $\sim$ 30 $h^{-1}$ Mpc to the west of the J1306 sightline is particularly extensive. Given that the J1306 sightline is highly transmissive, the proximity of these potential sources of ionizing photons raises the question of whether these nearby overdense regions play a significant role in ionizing the IGM in the vicinity of the quasar sightline. The recent measurement of the mean free path at $z=6.0$ by \citet{becker21} makes it possible to estimate what the mean free path should be at $z=5.7$. \citet{becker21} measure $\lambda_{\rm mfp} = 9.09^{+1.62}_{-1.28}$ proper Mpc at $z=5.1$, and $\lambda_{\rm mfp} = 0.75^{+0.65}_{-0.45}$ proper Mpc at $z=6.0$. Linearly interpolating between these two measurements, we find that the mean free path at $z=5.7$ should be approximately $\lambda_{\rm mfp} = 3.5$ proper Mpc, which corresponds to 16.4 $h^{-1}$ comoving Mpc. Referring to Figure \ref{fig:lae_map}, if $\lambda_{mfp} = 16.4$ $h^{-1}$ Mpc, then parts of the overdense regions in the J1306 field lie within a mean free path of the sightline. While this is a rough approximation, given that the mean free path will vary locally, it is at least plausible that these nearby overdense regions could contribute to the ionization state of the IGM in the vicinity of the quasar sightline. We also find that, for the simulated sightlines of \citet{nasir20}, highly transmissive, low-density sightlines are more likely to show an overdensity in adjacent radial bins in their surface density profile (similar to the J1306 field in Figure \ref{fig:sd_rad}) than their higher-opacity counterparts. For example, of the sightlines in the late reionization, short mean free path model, 55\% of the sightlines with $\tau_{\rm eff}^{50} \leq 3.0$ and normalized surface density $\leq 0.5$ within $R\leq 10$ $h^{-1}$ Mpc also had a normalized surface density of $\geq 1.25$ in either the $10-20$ or $20-30$ $h^{-1}$ Mpc bin, compared to 11\% of sightlines with $\tau_{\rm eff}^{50} \geq 5.0$. This trend holds for all three models of \citet{nasir20}, which suggests that adjacent overdensities may play a role in the high transmission of these sightlines. There is evidence from the literature that suggests LAEs may enhance the local photoionizating background. \citet{meyer19,meyer20} report an excess of \lya\ transmission spikes within $10-60$ cMpc from LAEs; this relationship between \lya\ flux and galaxy distance is additionally well-matched by the THESAN models \citep{garaldi22}. Similarly, \citet{kakiichi18} find that IGM \lya\ transmission is preferentially higher in the vicinity of luminous Lyman break galaxies, many of which also show \lya\ lines, and \citep{kashino23} find that IGM \lya\ transmission peaks 5 Mpc from [OIII] emitting galaxies at $5.7<z<6.14$. Given this observed link between galaxies and elevated \lya\ transmission, one possible interpretation of our observations of transmissive sightlines is that they can arise in less dense regions that are close enough to an overdensity to have an elevated ionizing background that contributes to its highly transmissive state. This interpretation is qualitatively consistent with both the galaxy UVB and ultra-late reionization scenarios.

\subsection{Opacity-density relation}\label{sec:sd-tau}
\begin{figure*}
\includegraphics[width=\textwidth]{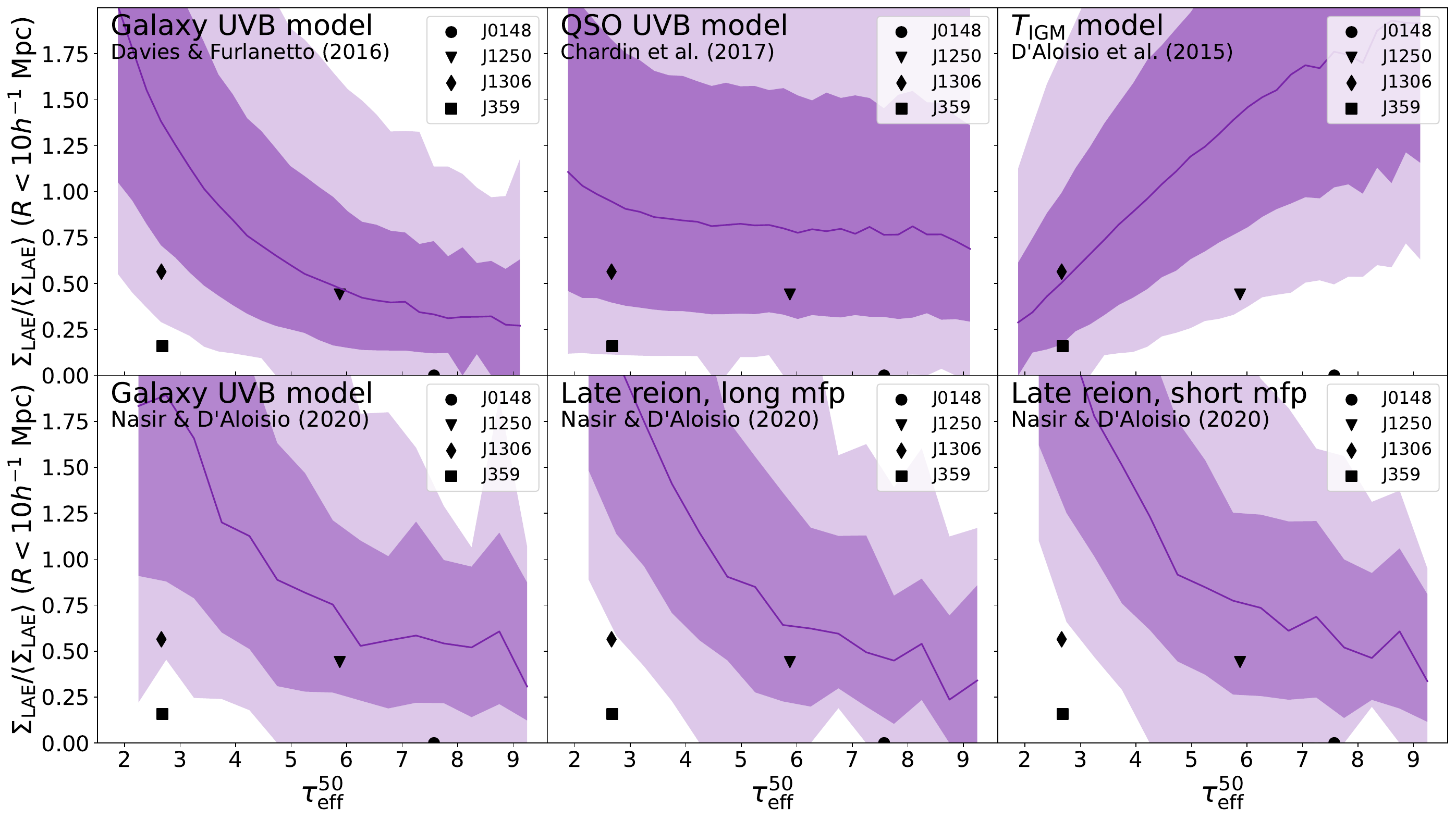}
\caption{Comparison of the measured surface density of LAE candidates within 10 $h^{-1}$ Mpc of the quasar sightline to model predictions for the relationship between opacity and LAE density. The models are the same as those in Figures \ref{fig:models-opaque} and \ref{fig:models-transmissive} and predictions are made using the full set of model sightlines spanning all opacity values. All surface densities are given normalized by the mean surface density measured over 15\arcmin\ $\leq\theta\leq $ 40\arcmin in each individual field.}
\label{fig:sd_tau_field}
\end{figure*}

Now that a number of extreme opacity QSO fields have been surveyed for LAEs, we can begin to characterize the relationship between \lya\ opacity and galaxy density at $z\sim5.7$. Figure \ref{fig:sd_tau_field} shows the measured surface density in the inner 10 $h^{-1}$ Mpc of all four fields as a function of the \lya\ effective opacity. Also shown are the predictions for the relationship between surface density of LAEs and \lya\ opacity in each of the models. These measurements are normalized by the mean surface density in their respective fields. 
 
The surface density measurements for transmissive sightlines put some pressure on fluctuating UVB and late reionization models, as the J359 measurement falls outside 98\% range for some of the model predictions. Further, we note that all four surface density measurements lie near or below the median predictions for all models. This outcome is unlikely to occur randomly; there is only a 6.25\% chance that four randomly drawn sightlines would lie below the median. The probability of reproducing our densities is as low as $<2$\%, moreover, given that some of the measurements lie below the 68\% and 98\% thresholds for the different models. This emphasizes the possibility that none of the models accurately capture the relationship between opacity and density across the full $\tau_{\rm eff}$ range.

\begin{figure*}
\includegraphics[width=\textwidth]{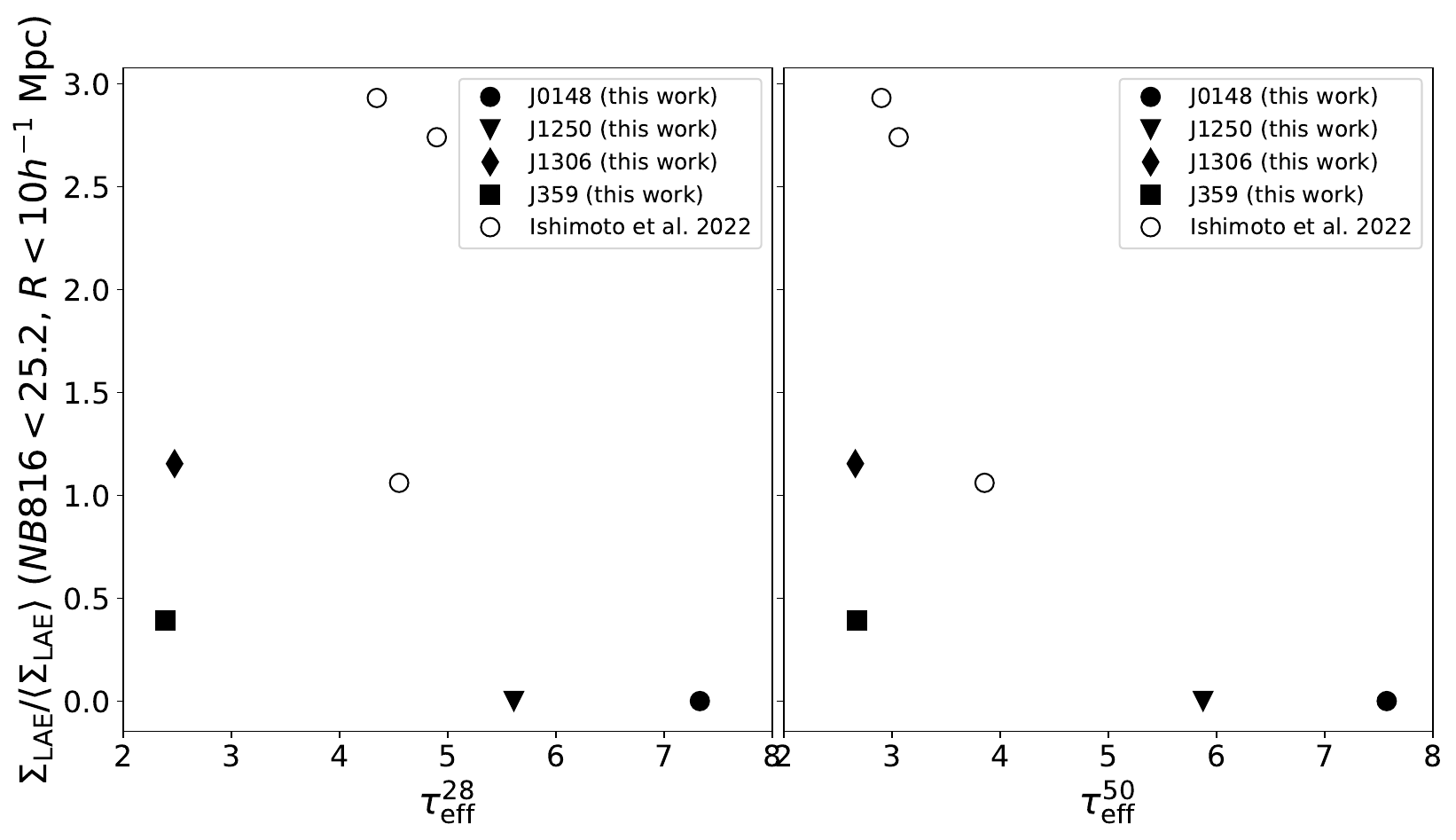}
\caption{Measured surface density of LAE candidates within 10 $h^{-1}$ Mpc of the quasar sightline as a function of $\tau_{\rm eff}^{28}$ (left) and $\tau_{\rm eff}^{50}$ (right). We include both $\tau_{\rm eff}$ windows here for comparison; for a discussion of the selection biases associated with each, see Section \ref{sec:sd-tau}. Included are all seven fields surveyed to date, presented in this work and \citet{ishimoto22}. For all seven fields, we match observational considerations as closely as possible, including the limiting magnitude, window of the $\tau_{\rm eff}$ measurement, and normalization. LAEs in all fields are selected down to the bright limit from \citet{ishimoto22} of NB816$\leq25.2$. Surface densities are given normalized by the mean surface density measured over 15\arcmin\ $\leq\theta\leq $ 40\arcmin in each field.}
\label{fig:sd_tau_allfields}
\end{figure*}

\begin{figure*}
\includegraphics[width=\textwidth]{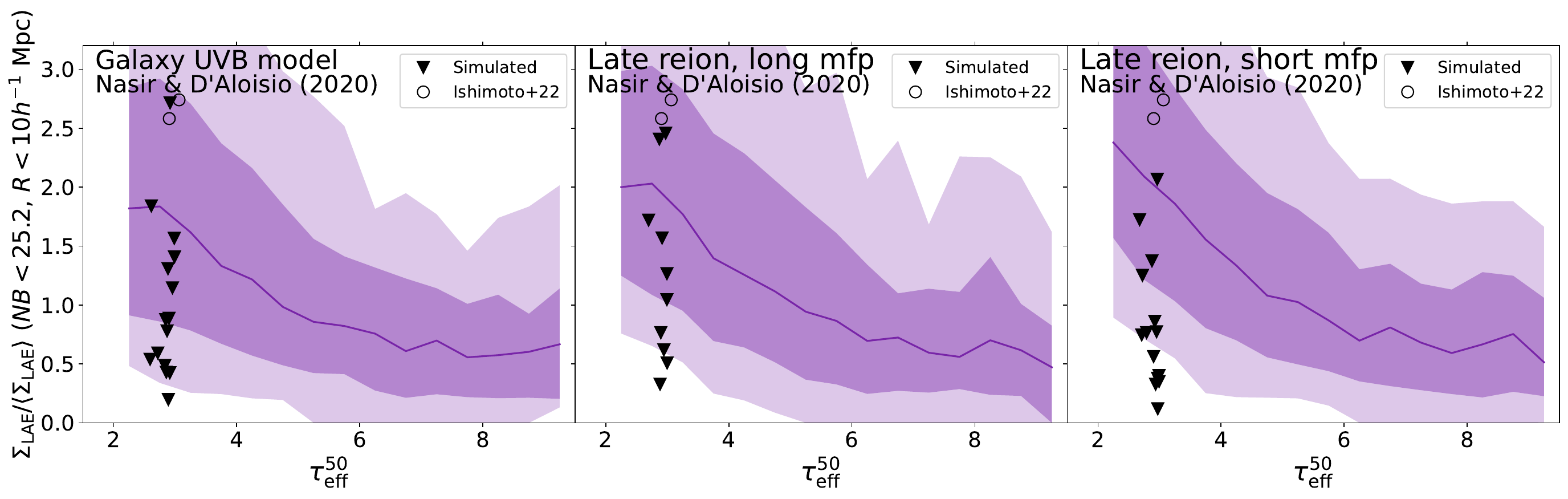}
\caption{Measured surface density of LAE candidates within 10 $h^{-1}$ Mpc of the quasar sightline as a function of $\tau_{\rm eff}^{50}$, for sightlines from \citet{ishimoto22} and simulated sightlines from the models of \citet{nasir20}. The simulated sightlines were selected to have $\tau_{\rm eff}^{28}\geq 4.0$ and $\tau_{\rm eff}^{50} \leq 3.0$, similar to the two overdense sightlines of \citet{ishimoto22}. The model predictions are made using $\tau_{\rm eff}^{50}$ values, and all surface densities are given normalized by the mean surface density measured over 15\arcmin\ $\leq\theta\leq $ 40\arcmin in each individual field. This figure illustrates that although these two lines of sight fall in the upper density range for their $\tau_{\rm eff}^{50}$ values, they are not consistent with simulated lines of sight from these models that were selected in the same way.}
\label{fig:selection_bias}
\end{figure*}

\begin{figure*}
\includegraphics[width=\textwidth]{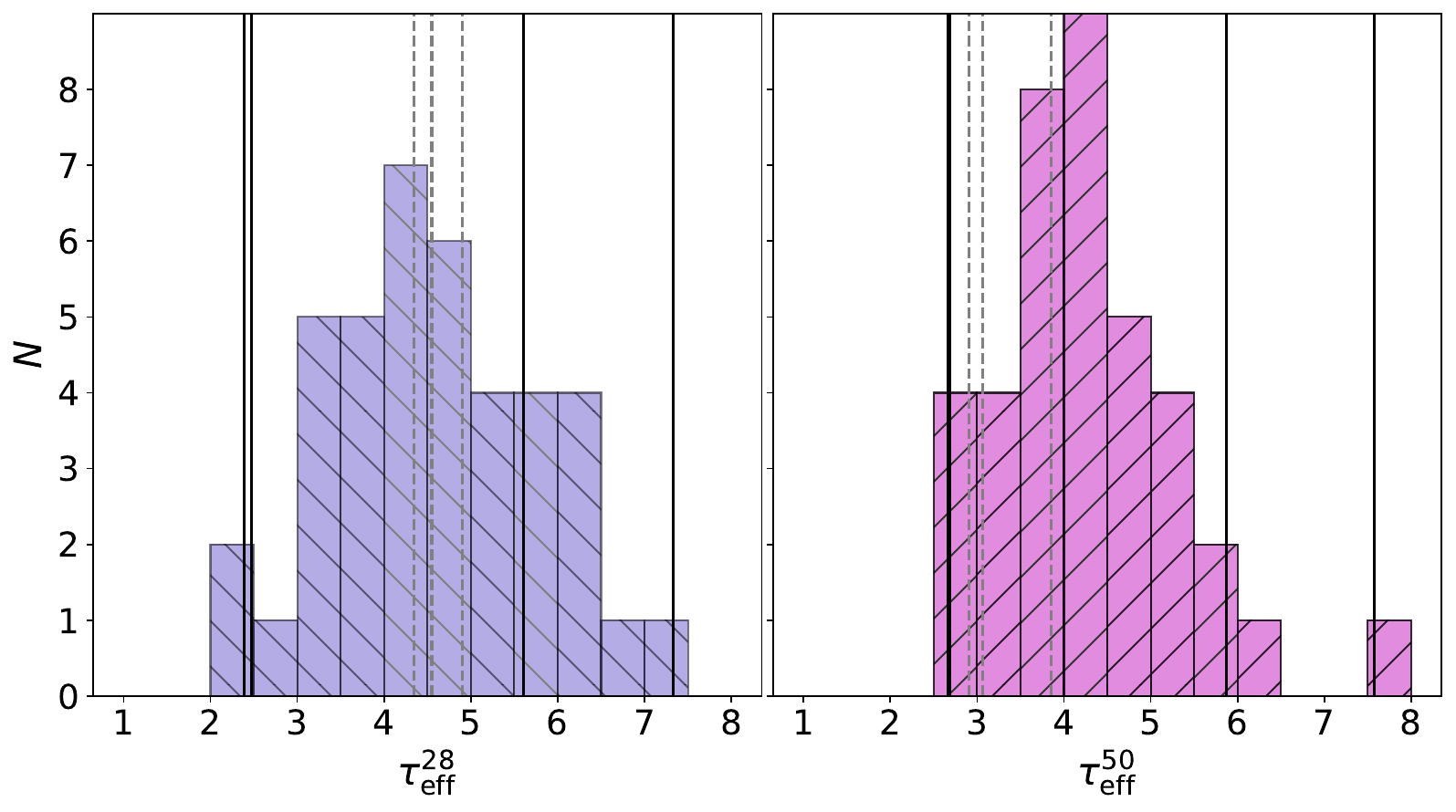}
\caption{Distribution of $\tau_{\rm eff}$ measurements for the quasar sample from \citet{zhu21}, measured both over 28 $h^{-1}$ Mpc (left) and 50 $h^{-1}$ Mpc (right). We show where the sightlines from this work and \citet{ishimoto22} fall in the distribution with solid black and dashed gray vertical lines, respectively. The sightlines from this work all fall in the wings of the global distribution, whereas the sightlines from \citet{ishimoto22} are more moderate.}
\label{fig:teff_dist}
\end{figure*}

\begin{figure*}
\includegraphics[width=\textwidth]{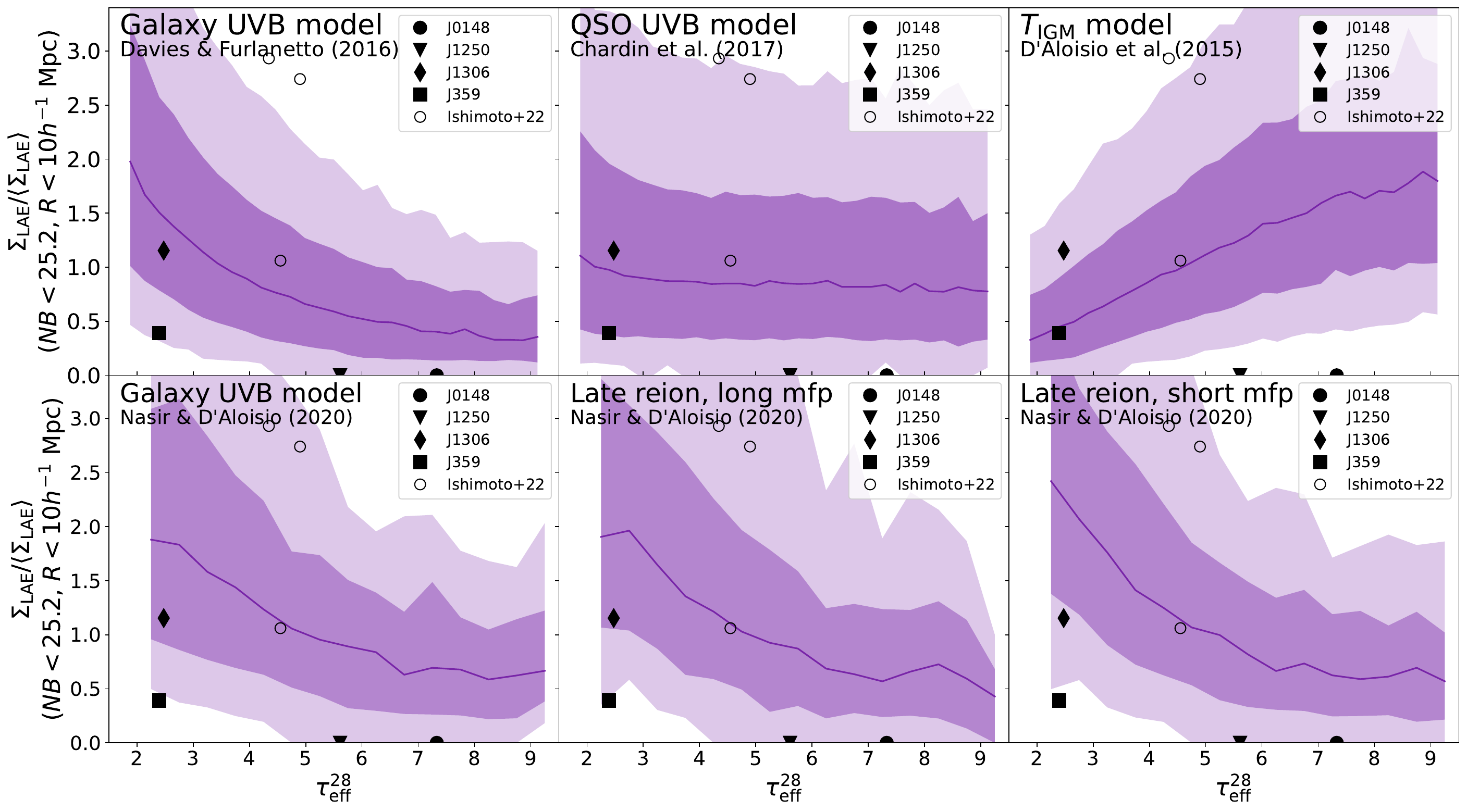}
\caption{Comparison of the measured surface density of LAE candidates within 10 $h^{-1}$ Mpc of the quasar sightline to model predictions for the relationship between opacity and LAE density. The models are the same as those used in Figures \ref{fig:models-opaque},\ref{fig:models-transmissive}, and \ref{fig:sd_tau_field}. Both observations and model predictions use opacity measurements made over 28 $h^{-1}$ Mpc and the $NB816\leq25.2$ magnitude limit of \citet{ishimoto22}. All surface densities are given normalized by the mean surface density measured over 15\arcmin\ $\leq\theta\leq $ 40\arcmin in each individual field.}
\label{fig:sd_tau_nb}
\end{figure*}

In addition to the four fields presented in this work, three additional fields have been surveyed by \citet{ishimoto22}. Their fields have $\tau_{\rm eff}$ values measured over 50 $h^{-1}$ Mpc of 4.17$\pm0.25$, 2.85$\pm0.04$, and 2.91$\pm0.03$, where these values are re-measured here from spectra reduced with a custom pipeline optimized for high-redshift QSOs (see Appendix \ref{sec:our-taus}).  Of these fields, the two with transmissive $\tau_{\rm eff}^{50}$ values are overdense, and the one with moderate $\tau_{\rm eff}^{50}$ is underdense in the vicinity of the quasar sightline. The $\tau_{\rm eff}$ values for all seven fields are summarized in Table 1.

In Figure \ref{fig:sd_tau_allfields}, we show the surface density in the inner 10 $h^{-1}$ Mpc of all seven quasar sightlines as a function of their $\tau_{\rm eff}^{28}$ (left panel) and $\tau_{\rm eff}^{50}$ (right panel). We use the bright limit from \citet{ishimoto22} of NB816$\leq25.2$ for all fields. The surface density measurement in each field is normalized by the mean surface density in that field, measured over $15\leq \theta \leq 40$ arcmin, as is done elsewhere in this work. 

In principle, the seven combined fields from this work and \citet{ishimoto22} present an opportunity to evaluate the opacity-density relation with greater sampling of the $\tau_{\rm eff}$ distribution.  At face value, low-$\tau_{\rm eff}$ lines of sight with high densities would support the fluctuating UVB and late reionization models. In practice, however, directly comparing these fields presents challenges. Field-to-field variations in depth and seeing and differences in methods for photometry, LAE selection, and completeness corrections all complicate the comparison (see \citep{ishimoto22} for an overview of their methodology). Further, the two sets of sightlines presented in this work and in \citet{ishimoto22} were selected in different manners. Our two highly opaque sightlines, J0148 and J1250, were selected based on the presence of long \lya\ troughs of 110 $h^{-1}$ and 81 $h^{-1}$ Mpc respectively.
The J1306 and J359 sightlines were selected based on their $\tau_{\rm eff}^{50}$ values, although the J1306 sightline was known to be transmissive over a longer segment of the \lya\ forest (eg \citealt{becker15}). In contrast, \citet{ishimoto22} selected their fields based on \lya\ forest opacities over the wavelength range of the NB816 filter, which corresponds to $\sim$28 $h^{-1}$ Mpc.  For a comparison of the $\tau_{\rm eff}$ measurements over 28 and 50 $h^{-1}$ Mpc windows, see Table 1. Our sightlines have similar $\tau_{\rm eff}$ values over these windows. Two of the three sightlines from \citet{ishimoto22}, however, show significant differences in their 28 or 50 $h^{-1}$ Mpc opacities.  In these cases, the forest is highly opaque over the 28 $h^{-1}$ Mpc window but shows strong transmission just outside it, giving a lower 50 $h^{-1}$ Mpc opacity. Because the sightlines from \citet{ishimoto22} were selected to be opaque over 28 $h^{-1}$ Mpc, they may not be representative of all sightlines with low $\tau_{\rm eff}^{50}$. Figure \ref{fig:sd_tau_allfields} illustrates the strong impact that the $\tau_{\rm eff}$ measurement window has on the results.

To understand the biases associated with selecting sightlines over the $NB816$ window, we investigated similar sightlines in the fluctuating UVB and late reionization models from \citet{nasir20}. We selected sightlines with $\tau_{\rm eff}^{28}\geq4.0$ and $\tau_{\rm eff}^{50}\leq 3.0$, similar to the sightlines from \citet{ishimoto22}. Of the 4000 simulated sightlines for each model, there are $10-15$ sightlines that meet these criteria. Similar to the real sightlines, the simulated ones uniformly show a strongly absorbed dark gap over the $NB816$ filter range, and strong transmission spikes over the remainder of the 50 $h^{-1}$ Mpc window. The densities of these sightlines sample the full range of density scatter shown in model predictions for density as a function of $\tau_{\rm eff}^{28}$. However, they are not representative of the density distribution for sightlines that are transmissive over 50 $h^{-1}$ Mpc. Figure \ref{fig:selection_bias} shows the surface density of these simulated sightlines and the sightlines observed by \citet{ishimoto22} compared to model predictions made over 50 $h^{-1}$ Mpc. Simulated sightlines that are selected to be opaque over 28 $h^{-1}$ Mpc are biased towards being underdense for their 50 $h^{-1}$ Mpc opacities. The two \citet{ishimoto22} sightlines with these opacity characteristics are denser than any of the simulated sightlines that were selected in the same manner.  
These sightlines are therefore also not obviously consistent with either the UVB or late reionization models.

We suggest that the $\tau_{\rm eff}^{28}$ window may be least impacted by selection effects because it reflects the selection criteria of \citet{ishimoto22} and because the $\tau_{\rm eff}$ measurements for the sightlines presented in this work are fairly consistent over both windows. At the limit of $NB816 < 25.5$ used in Figure\ref{fig:sd_tau_allfields}, highly opaque sightlines (on scales of 28 $h^{-1}$ Mpc) are correlated with galaxy underdensities, while the transmissive sightlines are mildly over- or underdense. Sightlines with moderate opacity, meanwhile, show a large scatter in observed density. Overall, three of the seven sightlines surveyed are underdense within 10 $h^{-1}$ Mpc of the quasar sightline and two are of average density. Although the overall sample tends towards lower densities, we note that most of these sightlines are selected to be atypical in terms of their $\tau_{\rm eff}$ values, and we do not expect them to average to unity as we would a larger, more representative sample. Figure \ref{fig:teff_dist} shows the distribution of $\tau_{\rm eff}$ values for the quasar sample of \citet{zhu21} measured over both 28 (left) and 50 (right) $h^{-1}$ Mpc windows. The opacity values for the sightlines discussed in this work are marked with vertical lines. The four sightlines presented in this work, which are mostly underdense, fall at the extreme ends of the distribution. The sightlines from \citet{ishimoto22}, which show a range of densities, fall in the center of the global distribution and are likely to be more representative of the majority of quasar sightlines at this redshift. 

Figure \ref{fig:sd_tau_nb} shows the surface density within 10 $h^{-1}$ Mpc of the quasar sightline as a function of $\tau_{\rm eff}^{28}$ for all sightlines from this work and \citet{ishimoto22}, compared to predictions from the models of \citet{nasir20}. Both the data and models use $\tau_{\rm eff}^{28}$ opacity measurements and the $NB816\leq25.2$ magnitude limit of \citet{ishimoto22}. We use $\tau_{\rm eff}^{28}$ values for this model comparison because they may be less impacted by selection effects than $\tau_{\rm eff}^{50}$ values, as discussed above. 

Altogether, these observations are not clearly consistent with any of the models considered here. The association of highly opaque sightlines and galaxy underdensities is explained well by fluctuating UVB and late reionization models, but these models do not obviously work well for the transmissive sightlines. On the other hand, the temperature model is in good agreement with the transmissive sightlines.

We can speculate on what may be happening at the low-opacity end. In a post-reionization IGM with a more homogeneous UVB, we expect that opacity will positively correlate with density.  This correlation may even be enhanced by temperature fluctuations for some period following reionization, as in the fluctuating temperature model. A homogeneous UVB is not expected at $z=5.7$; indeed, there is strong observational evidence for UVB fluctuations persisting as late as $z\sim5.3$ \citep{becker15,bosman18,eilers18,yang20,bosman21,zhu21}.
If the UVB is not as highly suppressed in underdense regions as the models considered here suggest, however, then these regions may quickly transition from being highly opaque to being transmissive once they are fully reionized, an evolution first suggested by \citet{keating20b}. 

A caveat of this work is the assumption that LAEs are a good tracer of the underlying density field, an assumption that is complicated near the end of reionization by how susceptible \lya\ photons are to attenuation by neutral gas. \citet{davies18} found that LAE surveys were $\sim$90\% likely to distinguish between fluctuating UVB and temperature models. However, there is some observational evidence, albeit at lower redshift, that LAEs either avoid some high-density peaks \citep{francis04,kashikawa07,huang22}, or tend to prefer lower-density regions \citep{cooke13}, possibly because higher-density regions have a stronger UVB that suppresses star formation \citep{kashikawa07,bruns12}. \citet{kashino20} surveyed Lyman break galaxies (LBGs) in the J0148 field and found an underdensity in the vicinity of the quasar sightline, which indicates that the J0148 underdensity is not the result of \lya\ suppression by neutral gas. However, it is unclear whether LBGs and LAEs in this field trace the same large scale structures, in part due to the broader redshift range spanned by the LBG selection ($\Delta z \sim 0.4$). It is also unclear whether a survey of LBGs in a field surrounding a transmissive sightline would similarly show the same density profile as the LAE population. A promising avenue for future work is therefore to consider other types of galaxy surveys to corroborate the results of the LAE selections. In addition to LBGs, sub-mm sureys, which probe massive, obscured galaxies, may be a useful probe of the density at these redshifts; \citet{qiong23} recently surveyed sub-mm galaxies in the J0148 field and reported an overdensity, although without redshifts it is unclear whether they are in proximity to the \lya\ trough.
It is also now possible to select galaxies at these redshifts based on their [OIII]$\lambda\lambda$4960,5008 emission with JWST/NIRCam, as done by the EIGER team \citep{kashino23}.

\section{Summary}\label{sec:summary}
We present an initial characterization of the relationship between IGM \lya\ opacity and galaxy density at $z=5.7$ by surveying Lyman-$\alpha$ emitting galaxies in the fields surrounding quasar sightlines with extreme values of \lya\ opacity. 
The relationship between IGM opacity and galaxy density on large ($\gtrsim$ 10 $h^{-1}$\ Mpc) transverse scales serves as a test of reionization models that predict the observed scatter in \lya\ opacity. Surveying sightlines over a wide range of \lya\ opacity, particularly extreme values, is necessary to characterize this relationship. We present two new surveys of LAEs towards the $z=6.02$ quasar SDSS J1306+0356 and the $z=6.17$ quasar PSO J359-06, whose sightlines show very low effective \lya\ opacity over 50 $h^{-1}$ Mpc along the line of sight ($\tau_{\rm eff}^{50} = 2.6$ and $\tau_{\rm eff}^{50} = 2.7$ for the J1306 and J359 fields respectively). We also re-select LAEs in the fields surrounding two highly opaque sightlines, towards ULAS J0148+0600 and SDSS J1250+3130, using the aperture photometry adopted for this work. 

We report an underdensity of LAEs within 10 $h^{-1}$ Mpc of both transmissive quasar sightlines. The results towards highly opaque sightlines are unchanged from previous works \citep{becker18,christenson21}; we find strong underdensities in the vicinity of both quasar sightlines. We note that the underdensities associated with \lya\ troughs span greater lateral extent than those associated with transmissive sightlines ($\gtrsim$ 20 $h^{-1}$ Mpc; see Section \ref{sec:environments}). We compare the measured surface density as a function of radius to predictions made by three broad types of models in Figure \ref{fig:models-transmissive}: fluctuating UVB models \citep{davies18,nasir20,chardin15,chardin17}, fluctuating temperature models \citep{daloisio15}, and ultra-late reionization models (\citealt{nasir20}, see also \citealt{kulkarni19,keating20a}. The correlation between highly opaque sightlines and galaxy underdensities strongly disfavors the temperature model, and the fluctuating UVB and late reionization models are unlikely to produce transmissive sightlines as underdense as those we observe.  None of the models, on their own, cleanly predict our observations of all four sightlines. 

 Our measurements allow us to begin characterizing the observed LAE surface density as a function of \lya\ effective opacity (see Figure \ref{fig:sd_tau_field}). The highly transmissive sightlines are sufficiently underdense within 10 $h^{-1}$ Mpc of the quasar sightline to be challenging for galaxy-driven UVB and late reionization models, which favor overdense regions associated with transmissive sightlines. Further, all of our observations fall below the median model predictions for the opacity-density relation, which hints that the models may not fully capture the physical conditions leading to sightlines with extreme opacity.

A total of seven fields surrounding quasar sightlines have now been surveyed in this manner. We show the LAE surface density as a function of \lya\ effective opacity of our four fields together with three from \citet{ishimoto22}  (Figure \ref{fig:sd_tau_allfields}). While the sightlines with extreme opacity are correlated to galaxy underdensities within 10 $h^{-1}$ Mpc of the quasar sightline, the sightlines of moderate opacity range from median density to significantly overdense. The association of highly opaque sightlines with galaxy underdensities is well-predicted by UVB and late reionization models. The association of highly transmissive sightlines with galaxy underdensities, however, is in possible tension with these models. One possible interpretation of these observations is that as reionization ends, the UVB transitions to a more homogeneous state more quickly than in the models considered here, causing the hot, recently reionized voids to rapidly become highly transmissive.  This evolution in the transmission of the voids was first suggested by \citet{keating20b}.

Further galaxy surveys, particularly towards transmissive sightlines, are needed for a more robust characterization of the relationship between opacity and density. If these further observations confirm the correlation between transmissive sightlines and galaxy underdensities, it would indicate that current reionization models do not adequately capture the ionizing sources and/or the sinks near the end of reionization. 

\acknowledgments
H.C. is supported by the National Science Foundation Graduate Research Fellowship Program under Grant No. DGE-1326120. H.C., G.B. and Y.Z. are supported by the National Science Foundation through grant AST-1615814 and AST-1751404. S.R.F. is supported by the National Science Foundation through awards AST-1812458 and AST-2205900 and by NASA through award 80NSSC22K0818. 

This research is based in part on data collected at Subaru Telescope, which is operated by the National Astronomical Observatory of Japan. We are honored and grateful for the opportunity of observing the Universe from Maunakea, which has cultural, historical and natural significance in Hawaii. The Hyper Suprime-Cam (HSC) collaboration includes the astronomical communities of Japan and Taiwan, and Princeton University. The HSC instrumentation and software were developed by the National Astronomical Observatory of Japan (NAOJ), the Kavli Institute for the Physics and Mathematics of the Universe (Kavli IPMU), the University of Tokyo, the High Energy Accelerator Research Organization (KEK), the Academia Sinica Institute for Astronomy and Astrophysics in Taiwan (ASIAA), and Princeton University. Funding was contributed by the FIRST program from Japanese Cabinet Office, the Ministry of Education, Culture, Sports, Science and Technology (MEXT), the Japan Society for the Promotion of Science (JSPS),
Japan Science and Technology Agency (JST), the Toray Science Foundation, NAOJ, Kavli IPMU, KEK, ASIAA, and Princeton University.

This research is based in part on observations collected at the European Organisation for Astronomical Research in the Southern Hemisphere under ESO programmes 084.A-0390, and 1103.A-0817.

The Pan-STARRS1 Surveys (PS1) have been made possible through contributions of the Institute for Astronomy, the University of Hawaii, the Pan-STARRS Project Office, the Max-Planck Society and its participating institutes, the Max Planck Institute for Astronomy, Heidelberg and the Max Planck Institute for Extraterrestrial Physics, Garching, The Johns Hopkins University, Durham University, the University of Edinburgh, Queens University Belfast, the Harvard-Smithsonian Center for Astrophysics, the Las Cumbres Observatory Global Telescope Network Incorporated, the National Central University of Taiwan, the Space Telescope Science Institute, the National Aeronautics and Space Administration under Grant No. NNX08AR22G issued through the Planetary Science Division of the NASA Science Mission Directorate, the National Science Foundation under Grant No. AST-1238877, the University of Maryland, and Eotvos Lorand University (ELTE), the Los Alamos National Laboratory, and the Gordon and Betty Moore Foundation.

This paper makes use of software developed for the Large Synoptic Survey Telescope. We thank the LSST Project for making their code available as free software at http://dm.lsst.org.

This research made use of the following additional software: Astropy \citep{astropy2013, astropy2018}, matplotlib \citep{hunter2007}, and scipy\citep{scipy2020}, including numpy \citep{numpy2011}.

The authors wish to recognize and acknowledge the very significant cultural role and reverence that the summit of Mauna Kea has always had within the indigenous Hawaiian community. We are most fortunate to have the opportunity to conduct observations from this mountain.

\bibliographystyle{apj.bst}
\bibliography{references.bib}

\appendix 
\section{Lyman-alpha Opacity of Quasar Sightlines}\label{sec:phot-opacity}
Following \citet{becker18,christenson21}, we use our imaging data to estimate the \lya\ effective opacity for the highly transmissive J1306 and J359 sightlines. Measurements made from the imaging data are comparable to spectroscopic measurements made over 28 $h^{-1}$ Mpc centered on the NB0816 filter wavelengths, and represents an effective opacity measurement made over the width of the NB816 filter. The general procedure is as follows: for each quasar, we begin by measuring the NB816 and HSC-$i2$ fluxes from the imaging data following Section \ref{sec:photometry}. We then convolve each quasar spectrum with the $i2$ transmission curve and scale them so that the transmission-weighted mean flux matches the $i2$ flux measured from the imaging data. We then estimate the unabsorbed continuum flux expected at the \lya\ wavelength from PCA fits for the blue-side continuum of each quasar spectrum. Combining these measurements, we calculate the effective opacity as $\tau_{\rm{eff}} = -\rm{ln}(F_{\lambda}^{NB816}/F_{\lambda}^{\rm{cont}})$.

For the J1306 sightline, we measure $F_{\lambda}^{NB816}= (11.2\pm 0.2)\times10^{-19}$ erg s$^{-1}$ cm$^{-2}$ \AA$^{-1}$ and $F_{\lambda}^{i2}= (28.3\pm 0.2)\times10^{-19}$ erg s$^{-1}$ cm$^{-2}$ \AA$^{-1}$, and estimate that the unabsorbed continuum is $F_{\lambda}^{\rm{cont}}\simeq 1.6\times10^{17}$ erg s$^{-1}$ cm$^{-2}$ \AA$^{-1}$. We therefore measure $\tau_{\rm eff} = 2.64\pm0.02$. For comparison, we measure $\tau_{\rm eff}^{28}= 2.475\pm0.010$ from the X-Shooter spectrum. The uncertainty in $\tau_{\rm eff}$ is based on the propagated uncertainty in $F_{\lambda}^{NB816}$ and does not account for uncertainty in the estimated continuum. 

For the J359 sightline, we measure $F_{\lambda}^{NB816}= (9.7\pm 0.2)\times10^{-19}$ erg s$^{-1}$ cm$^{-2}$ \AA$^{-1}$ and $F_{\lambda}^{i2}= (7.8\pm 0.1)\times10^{-19}$ erg s$^{-1}$ cm$^{-2}$ \AA$^{-1}$, and estimate that the unabsorbed continuum is $F_{\lambda}^{\rm{cont}}=0.8\times10^{17}$ erg s$^{-1}$ cm$^{-2}$ \AA$^{-1}$. We therefore measure $\tau_{\rm eff} = 2.26\pm0.02$.  From the spectra, we measure $\tau_{\rm eff}^{28} = 2.338\pm0.01$ over the filter width. For both sightlines, if we assume a 20\% uncertainty in the continuum, the uncertainty in our measurements from the imaging becomes $\pm 0.09$.

\section{\lya\ Opacity Measurements for Ishimoto et al. (2022) Lines of Sight}\label{sec:our-taus}

\LongTables
\begin{deluxetable}{lccccc}
\tablewidth{0pc}
\tablecaption{Effective opacity measurements for QSO sightlines \citet{ishimoto22}}\label{tab:taus_comparison}
\tabletypesize{\scriptsize}
\tablehead{QSO  & $\tau_{\rm eff}^{50,a}$ (this work) & $\tau_{\rm eff}^{50,b}$ (\citet{ishimoto22}) & $\tau_{\rm eff}^{28,c}$ (this work) & $\tau_{\rm eff}^{28,d}$ (\citet{ishimoto22})}
\startdata
SDSS J1137+3549  & $2.904\pm0.042$ & $3.07\pm0.03$ & $4.344\pm0.227$ & $5.58\pm0.62$\\
SDSS J1602+4228  & $3.063\pm0.038$ & $3.23\pm0.05$ & $4.898\pm0.308$ & $6.05\pm0.91$\\ 
SDSS J1630+4012 & $3.857\pm0.184$ & $5.47\pm0.86$ & $4.550\pm0.477$ & $>$5.06$^e$
\enddata
\tablenotetext{a}{Effective opacity used in this work, measured over a 50 $h^{-1}$ Mpc window centered at 8177 \AA}
\tablenotetext{b}{Effective opacity from \citet{ishimoto22}, measured over a 50 $h^{-1}$ Mpc window centered at 8177 \AA}
\tablenotetext{c}{Effective opacity used in this work, measured over a 28 $h^{-1}$ Mpc window centered at 8177 \AA}
\tablenotetext{d}{Effective opacity from \citet{ishimoto22}, measured over a 30 $h^{-1}$ Mpc window centered at 8177 \AA}
\tablenotetext{e}{Lower limit}
\end{deluxetable}


~\label{tab:taus_comparison}

\begin{figure*}
\includegraphics[width=\textwidth]{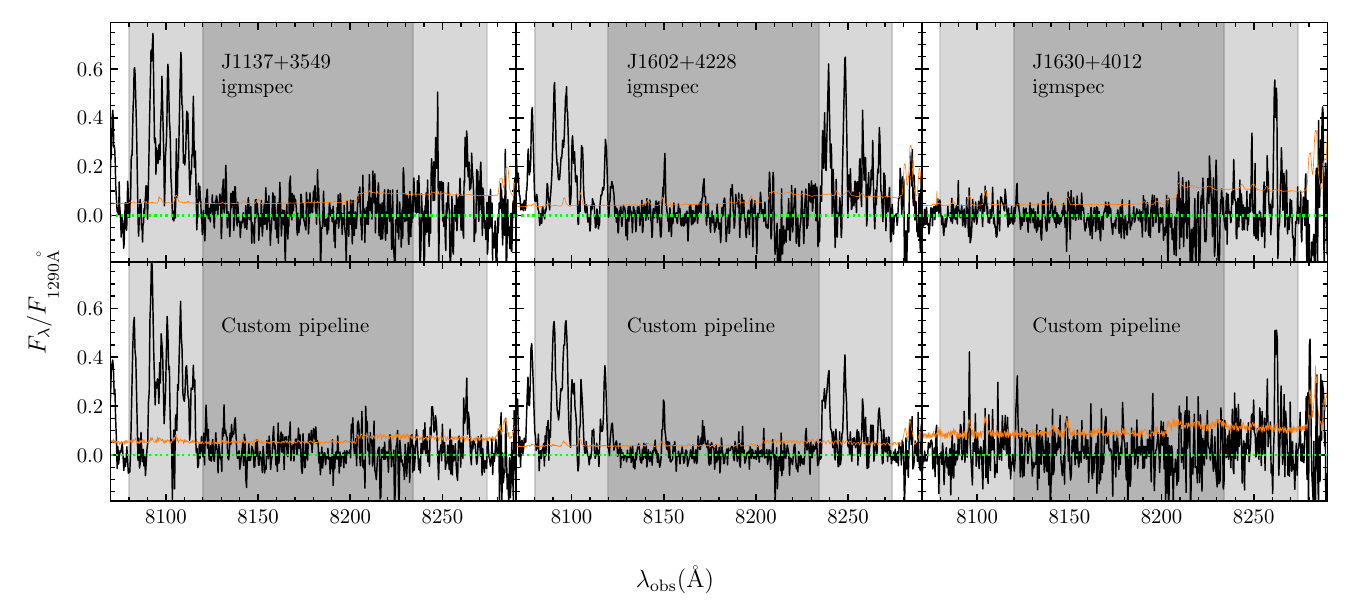}
\caption{Partial spectra of quasars J1137+3549, J1602+4228, and J1630+4012 (left to right). The top panels show the spectra for these objects used by \citet{ishimoto22}, which were selected from the {\tt igmspec} database \citep{prochaska17}.  The bottom panels show reductions using a pipeline customized for high-redshift QSOs \citet[e.g.,][]{becker19,zhu21,zhu22}. The solid orange lines indicate the flux error and the green dotted line marks a flux of zero. The darker gray shaded rectangles indicate the FWHM of the NB816 filter, which corresponds to 28 $h^{-1}$ Mpc and the lighter shaded regions indicate the 50 $h^{-1}$ Mpc interval over which effective opacity measurements are made. The opacity measurements made from these spectra, both in this work and in \citet{ishimoto22}, are summarized in Table 4.}
\label{fig:cmp_spectra}
\end{figure*}

In this work we use updated $\tau_{\rm eff}$ values for the three lines of sight included in \citet{ishimoto22}.  \citet{ishimoto22} used Keck ESI spectra from the publicly available {\tt igmspec} database \citep{prochaska17}.  In contrast, we use versions of these data reduced using a custom pipeline that has been highly optimized for high-redshift QSO spectra \citep[for a description of the pipeline, see][]{lopez16, becker19, zhu21}.  The custom reductions for all three were first presented in \cite{becker19}, while J1137 and J1602 were also presented in \citet{zhu21,zhu22}.  Our measurements of $\tau_{\rm eff}$ over the two wavelength regions indicated in Figure~\ref{fig:cmp_spectra}, corresponding to 28 and 50 $h^{-1}$ Mpc, are given in Table~\ref{tab:taus_comparison}, along with values from \citet{ishimoto22}.

We find somewhat lower values of $\tau_{\rm eff}^{28}$ for J1137 and J1602, and a lower $\tau_{\rm eff}^{50}$ for J1630.  For J1137, our reduction reveals transmission peaks near 8134 and 8180 \AA.  Taking the transmission from these peaks alone gives $\tau_{\rm eff}^{28} = 4.887 \pm 0.391$, which should be an upper limit on the effective opacity over the entire window as we are assuming that all other pixels have zero transmission.  This value is consistent with our measurement in Table~\ref{tab:taus_comparison}.  In the case of J1602, the higher $\tau_{\rm eff}^28$ value in \citet{ishimoto22} is explained by the presence of a spurious negative feature near 8212 \AA, which is not as strong in our reduction.  Similarly, the {\tt igmspec} reduction of J1630 appears to show a slight negative bias over 8160--8220 \AA, which helps to explain the difference in the $\tau_{\rm eff}^{50}$ values.

\section{Globally normalized LAE maps}\label{sec:globalnorm}
\begin{figure*}
\includegraphics[width=\textwidth]{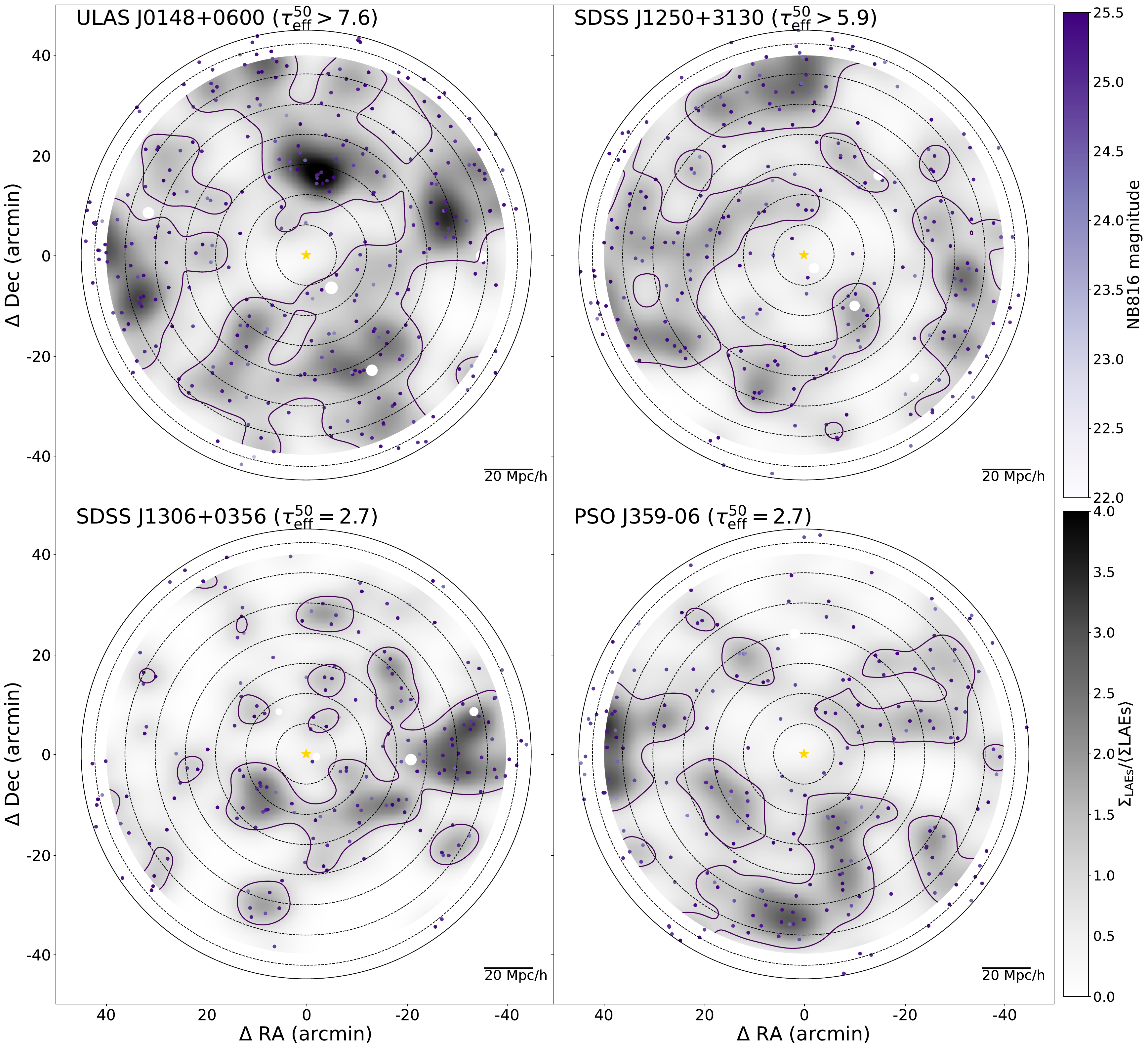}
\caption{Distribution of LAE candidates in all four fields: J0148 (top left), J1250 (top right), J1306 (bottom left), and J359 (bottom right). The LAE candidates are assigned a color that indicates their NB816 magnitude. The grayscale shading in the background indicates the surface density of LAE candidates, which we calculate by kernel density estimation and normalized by the global mean surface density of all four fields, measured over $15\arcmin \leq \Delta \theta \leq 40\arcmin$. This surface density is corrected for spatial variations in completeness as described in Section \ref{sec:completion}. The field is centered on the quasar position, which is marked with a gold star, and the concentric dotted rings indicate 10 $h^{-1}$ Mpc intervals from the quasar position. The solid ring marks the edge of the field, 45\arcmin\ from the quasar position. Portions masked out of the field in white are obscured by foreground stars.}
\label{fig:lae_map_global}
\end{figure*}

In Section \ref{sec:results}, we present maps of the LAE candidates in the J0148, J1250, J1306, and J359 fields. Those maps are normalized by the mean surface density in each field, calculated over $15\arcmin\leq \Delta \theta \leq 40\arcmin$. Normalizing the maps in this way allows us to self-consistently compare the vicinity of the quasar sightline to the rest of the field and determine whether the center of the field is over- or underdense relative to its surroundings. This type of normalization is also useful for making comparisons between fields, as it mitigates differences in depth, seeing, and other observational considerations, and it is the normalization used for all figures in the main body of the text.

However, we can also use the four fields we have observed to date to estimate a global mean surface density and compare the four fields on an absolute scale. This global normalization is limited by the small number of fields observed to date, but is useful for considering how the environments of these four sightlines compare to each other. Figure \ref{fig:lae_map_global} shows the LAE maps from Section \ref{sec:results}. As previously, the fields are centered on the quasar position (yellow star), foreground stars are masked out in white, and the concentric dotted rings indicate 10 $h^{-1}$ Mpc intervals. The purple shading indicates the NB816 magnitude of the LAEs, and the grayscale shading indicates the surface density of LAEs (see Section \ref{sec:results} for details of the calculation). The surface density is completeness corrected (see Section \ref{sec:completion}) and normalized by the global mean surface density, which we measure over the $15\arcmin\leq \theta \leq40\arcmin$ region of all four fields.

On an absolute scale, all four sightlines are underdense within 10 $h^{-1}$ Mpc of the quasar, consistent with the maps that are normalized individually (see Figure \ref{fig:lae_map}). However, we note that the J0148 field, while underdense in the vicinity of the quasar sightline, seems to reside in a higher density region overall than the other three fields.

\section{Comparison to Christenson et al. 2021 selections}\label{sec:c21}
\begin{figure*}
\includegraphics[width=0.48\textwidth]{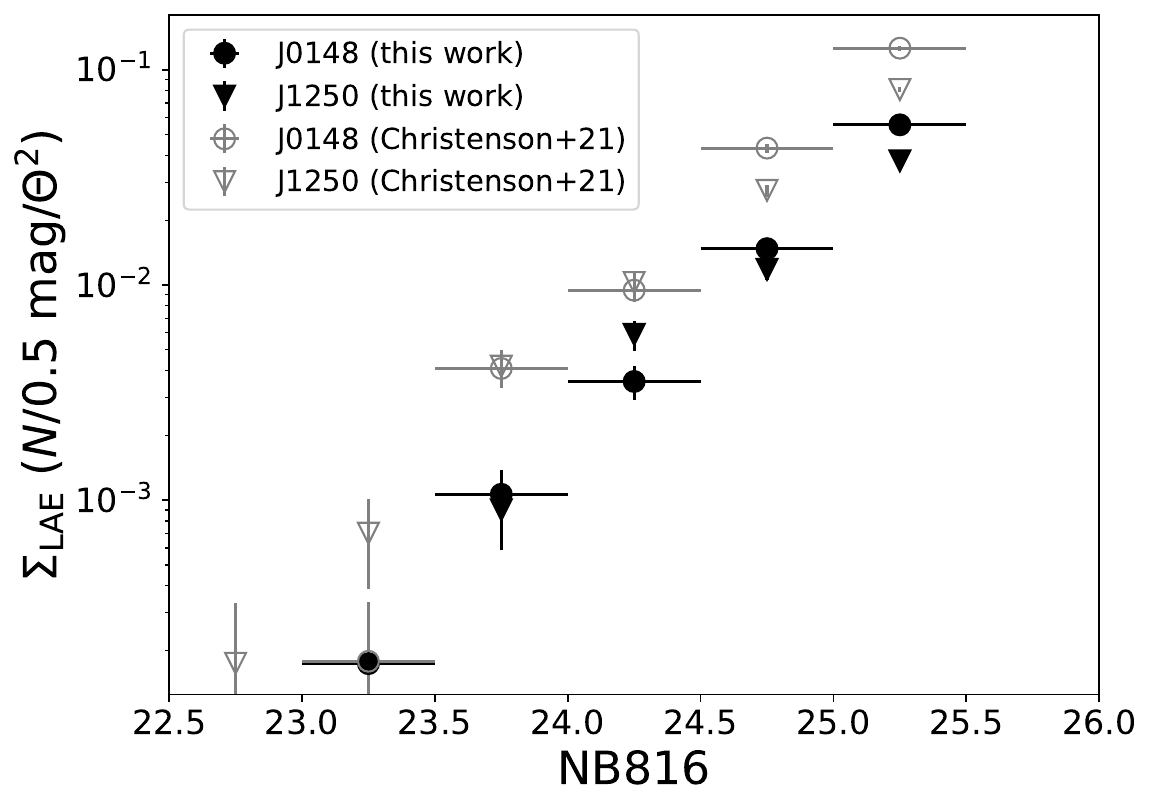}
\caption{Completeness-corrected surface density of LAE candidates in the J0148 and J1250 fields in this work (filled markers) and \citet{christenson21} (open markers).}
\label{fig:sd_mag_c21}
\end{figure*}


Here we compare the selections of LAEs made in the J0148 and J1250 fields in this work to those published in \citet{christenson21}. In this work, we select 298 LAEs in the J0148 field and 247 in the J1250 field, compared to 641 in the J0148 field and 428 in the J1250 field in \citet{christenson21}. We show the surface density as a function of $NB816$ magnitude for both selections in Figure \ref{fig:sd_mag_c21}, and refer the reader to Figures \ref{fig:lae_map}, \ref{fig:sd_rad}, and the corresponding figures in \citet{christenson21} for a comparison of the LAE maps and radial surface density distributions. There are two primary differences between these catalogs. First, as discussed in Section \ref{sec:photometry}, is the use of aperture fluxes as the primary photometric measurement in this work. Aperture fluxes are expected to have lower signal-to-noise compared to PSF fluxes and hence produce fewer detections. Second, we have made a more careful measurement of the seeing in this work, making use of bright stars selected to be bright, but not saturated, and the seeing tends to be slightly better than previously estimated. The PSF fluxes from \citet{christenson21}, which are fit to a broader profile than the true extent of the sources, therefore tend to be higher than the aperture fluxes.
For these two reasons, the measured surface density is not consistent between the two selections. We note that brightest objects consistently appear in both catalogs but tend to fall in fainter magnitude bins in this work, which is the reason for the poor agreement in the brighter magnitude bins. Despite these differences, the key results of \citet{christenson21} are unchanged in this work; both highly opaque sightlines display clear underdensities within 20 $h^{-1}$ Mpc of the quasar sightline, and the large-scale structure of the field is consistent between the two selections.

\end{document}